\documentclass[aps,prd,preprint,showpacs,superscriptaddress,tightenlines,nofootinbib]{revtex4-1}

\usepackage{fullpage}
\usepackage{graphics}
\usepackage{graphicx}
\usepackage{epsfig}
\usepackage{amsmath}
\usepackage{amssymb}
\usepackage{amsthm}
\usepackage{calc}
\usepackage{hyperref}
\hypersetup{
    bookmarks=true,%
    colorlinks,%
    citecolor=blue,%
    filecolor=blue,%
    linkcolor=blue,%
    urlcolor=blue
}
\usepackage{color}

\newcommand{\beq}{\begin{equation}}
\newcommand{\eeq}{\end{equation}}
\newcommand{\bea}{\begin{equation}\begin{aligned}}
\newcommand{\eea}{\end{aligned}\end{equation}}

\newcommand{\dd}{\text{d}}

\newcommand\undermat[2]{%
  \makebox[0pt][l]{$\smash{\underbrace{\phantom{%
    \begin{matrix}#2\end{matrix}}}_{\text{$#1$}}}$}#2}

\begin{document}
\title{Entropy of non-extremal STU black holes: The $F$-invariant unveiled}
\author{G\'abor S\'arosi}
\affiliation{Department of Theoretical Physics, Institute of
Physics, Budapest University of Technology, H-1521 Budapest,
Hungary\\ \vspace{0.2cm}}
\affiliation{Kavli Institute for Theoretical Physics, Kohn Hall,
University of California,
Santa Barbara CA 93106-4030, USA\\ \vspace{0.2cm}}
\date{August 26, 2015, last revised: \today}

\begin{abstract}
We find a manifestly U-duality invariant formula for the Bekenstein-Hawking entropy of the most general 4 dimensional, stationary, asymptotically flat, non-extremal STU black holes constructed recently by Chow and Comp\`ere. The expression is entirely in terms of asymptotic charges. It involves the "scalar charges" of the black hole which still need to be solved in terms of the dyonic charges and the mass. We discuss how the formula reduces to some of the known results as the Klauza-Klein black hole and the dilute gas limit of Cveti\v c and Larsen. We give the expected generalization to an $E_{7(7)}$ invariant in the case of maximal $\mathcal{N}=8$ supergravity.
\end{abstract}
\maketitle{}
\tableofcontents
\pagebreak

\section{Introduction}

Recently, Chow and Comp\`ere\cite{compere1,compere2} have constructed the most general asymptotically flat, rotating, non-extremal dyonic black hole solutions admitted by ungauged $\mathcal{N}=2$ supergravity coupled to three vector multiplets in four dimensions, or the so called STU model. 
The Bekenstein-Hawking entropy of these black holes can be written in the form
\beq
S=2\pi ( \sqrt{\Delta+F}+\sqrt{-J^2+F}),
\eeq
where $J$ is the angular momentum, $\Delta$ is Cayley's hyperdeterminant formed from the dyonic charges of the black hole and $F$ is a so far unidentified U-duality invariant given in terms of auxilary parameters.

 The main result of this paper is the formula
\bea
\label{eq:firstF}
F&=M^4+\frac{ M^2}{12}\text{Tr}  K_{12}-\frac{M }{24}\text{Tr} ( K_{12}R)\\
&+\frac{1}{192}\bigg( \text{Tr}(K_{11}^2)-\text{Tr}(K_{11}K_{22})-\frac{1}{2}(\text{Tr}R^2)^2+\text{Tr}(R^4)\bigg),
\eea
where $M$ is the mass of the black hole while $R$, $K_{11}$, $K_{22}$ and $K_{12}$ are certain $6\times 6$ matrices depending on six scalar charges, eight dyonic charges, and on the asymtotic values of the scalar fields as well. As we will explain, these matrices transform in the adjoint of the U-duality group $SL(2)^{\times 3}_U$ and therefore the expression for $F$ is manifestly U-duality invariant. We also give the $F$ invariant in the case of non-zero NUT charge in e.q. \eqref{eq:Finv1}. It is important to note that while the scalar charges are functions of the mass and the dyonic charges, the asymptotic values are independent parameters and therefore the general black hole entropy depends on the scalar hair. This confirms that there is no attractor mechanism for general non-extremal STU black holes\cite{kallosh,goldstein}. This observation can also be reinforced by the following argument. The dyonic charges, transforming under the U-duality group, form a prehomogeneous vector space\cite{satokimura}. This means that there is only one algebraically independent continous invariant, in this case Cayley's hyperdeterminant. Since the mass and the angular momentum are U-duality invariants\cite{andrianopoli2014extremal} the only way that the $F$-invariant can be non-trivial is that it depends on the scalar hair.

The construction of the building blocks for the $F$-invariant is most conveniently done using the toolbox of classifying certain entangled systems under stochastic local operations and classical communication (SLOCC) as three qubits\cite{cirac}, four qubits\cite{verstraete} and three fermions with six single particle states\cite{levvran1}. The idea of formulating black hole physics in supergravity in the language on quantum entanglement is not new. The entropy of extremal STU black holes is always expressed with Cayley's hyperdeterminant $\Delta$ which is a measure of tripartite entanglement for three qubits\cite{duff2,CKW}. The timelike reduced STU model can be described in the language of four qubit entanglement, where the line element is given by a quadratic entanglement measure\cite{bergshoeff2009generating,levay4qubit}. Moreover, extremal black holes with nilpotent charge vector can be classified in the language of four qubit entanglement as shown in \cite{duff1}. For a review of this so called black hole/qubit correspondence see \cite{levay}.

The organization of this paper is as follows. In sec. \ref{sec:conv}. we give a quick review of the conventions we use for STU supergravity and its reduction to the 3d coset model $SO(4,4)/SL(\mathbb{R})^{\times 4}$. In sec. \ref{sec:bhsolution}. we review the aspects of the black hole solution of Chow and Comp\`ere that we need. In sec. \ref{sec:actionofU}. we describe the action of the U-duality group $SL(2,\mathbb{R})^{\times 3}$ on the coset element putting particular emphasis on the fact that the   generators for U-duality live in a different splitting of $\mathfrak{so}(4,4)$ than the one defining the 3d coset model $SO(4,4)/SL(\mathbb{R})^{\times 4}$. After this, we describe how the 16 asymptotic charges of the black hole fit into the $9\oplus 8\oplus 8$ representation of the U-duality group. We give the relevant covariants that can be used to generate polynomial invariants of this representation space in the simple language of fermionic entanglement theory. In sec. \ref{sec:Finv}. we use these covariants as building blocks to find the announced expression for the $F$-invariant. Section \ref{app:generalscalars}. contains the formulas which are needed to compute the $F$-invariant dressed with the asymptotic values of the scalars. In section \ref{sec:scalcharg}. we give three constraints among the scalar and physical charges and speculate if they are enough to determine the scalar charges uniquely. As an illustration we solve the constraints for some special cases like the Klauza-Klein black hole and the dilute gas limit of \cite{larsen}. We find agreement with the literature. In sec. \ref{sec:generalize}. we describe how we expect our results to generalize to an $E_{7(7)}$ invariant for maximal $\mathcal{N}=8$ supergravity.

We include two appendices containing some explicit formulas which we found too cumbersome to include in the main text. In appendix \ref{sec:liealg}. we set up our conventions for the Lie-algebra of $\mathfrak{so}(4,4)$ and describe the three different ways of selecting an $\mathfrak{sl}_2^{\times 4}$ subalgebra with the remaining generators transforming in the fundamental $(2,2,2,2)$ representation. In appendix \ref{app:stumatrices}. we give the explicit forms of the matrices $R$, $K_{11}$, $K_{12}$ and $K_{22}$ appearing in expression \eqref{eq:firstF} for the $F$-invariant.

\section{$\mathcal{N}=2$ supergravity}
\label{sec:conv}

The term STU model refers to ungauged $\mathcal{N}=2$ supergravity in four dimensions coupled to 3 vector multiplets\cite{cremmer,duff4}. The model is of central importance in string theory as it can be obtained from most string theories and is related to various other supergravity theories through dualities. Also, a suitable solution of the STU model can be used\cite{cvetic} to generate all single centered stationary black holes of maximal $\mathcal{N}=8$ supergravity\cite{julia1,julia2} which describes the low energy limit of M-theory compactified on $T^7$. The black hole solutions in the theory can be obtained from the bosonic part of the action without the hypermultiplets. There are several formulations of this action, here we simply pick one and refer to the literature on details. The action we choose is
\beq
\label{eq:4daction}
S=\frac{1}{16 \pi}\int \dd x^4 \sqrt{|g|}\lbrace -\frac{R}{2}+G_{i\bar j}\partial_\mu \tau^i \partial^\mu \bar \tau^{\bar i}+(\text{Im} \mathcal{N}_{IJ}(\mathcal{F}^I)_{\mu\nu}(\mathcal{F}^J)^{\mu\nu}+\text{Re} \mathcal{N}_{IJ}(\mathcal{F}^I)_{\mu\nu}(\star \mathcal{F}^J)^{\mu\nu})\rbrace.
\eeq
Here, $\mathcal{F}^I$, $I=0,1,2,3$ are four $U(1)$ gauge field strengths and $\star \mathcal{F}^I$ is the dual of $\mathcal{F}^I$. The complex scalars $\tau^i$, $i=1,2,3$ are coordinates of the projective special K\"ahler manifold $[SL(2,\mathbb{R})/SO(2)]^{\times 3}$ and $G_{i\bar j}$ is the K\"ahler metric of this space. The matrix $\mathcal{N}_{IJ}$ is a certain $4\times 4$ matrix depending only on the scalars $\tau^i$. As we do not need them in this paper, for explicit forms of $G_{i\bar j}$ and $\mathcal{N}_{IJ}$ we refer to the literature\cite{levay4qubit,compere2}. We will usualy use the decomposition of $\tau^i$ into real in imaginary parts as 
\beq
\tau^i=x_i+i y_i.
\eeq
To describe stationary black hole solutions we proceed with the usual procedure of dimensional reduction along the time coordinate. For details of this procedure see e.g. \cite{breitenlohner,compere2}. The ansatz for the metric and the gauge fields is
\bea
\dd s^2 &= -e^{2 U} (\dd t+\omega)^2+e^{-2U} \dd s^2_{(3d)},\\
\mathcal{F}^I &=\dd \mathcal{A}^I=\dd(\zeta^I(\dd t+\omega)+A^I).
\eea
Here, $\dd s^2_{(3d)}=h_{ab}\dd x^a \dd x^b$ is the three dimensional line element with $a,b$ being 3d indices. The scalars $U$ and $\zeta^I$ and the one-forms $\omega$ and $A^I$ are considered to be three dimensional fields. One then dualizes $\omega$ and $A^I$ to scalars $\sigma$ and $\tilde \zeta_I$ as
\bea
\dd \tilde \zeta_I &=\text{Re}\mathcal{N}_{IJ} \dd \zeta^J-e^{2 U}\text{Im}\mathcal{N}_{IJ}  \star_{(3d)}(\dd A^J+\zeta^J \dd \omega),\\
\dd \sigma &= e^{4 U} \star_{(3d)} \dd \omega + \zeta^I \dd \tilde \zeta_I-\tilde \zeta_I \dd \zeta^I.
\eea
The resulting three dimensional theory is Euclidean gravity coupled to 16 real scalars $\lbrace U,\sigma,\tau^i,\bar \tau^{\bar i},\zeta^I,\tilde \zeta_I \rbrace$ parametrizing the coset space $SO(4,4)/SL(2,\mathbb{R})^{\times 3}$. The Lagrangian can be written as
\beq
\label{eq:3daction}
\mathcal{L}=-\frac{1}{2}\sqrt{h}R[h]+\sqrt{h}g_{mn}\partial_a \Phi^m \partial^a \Phi^n,
\eeq
where $\Phi^m$ denotes the scalars with $m=1,..,16$. The metric $g_{mn}$ on $SO(4,4)/SL(2,\mathbb{R})^{\times 4}$ is expressed through the line element as
\bea
\label{eq:cosetmetric}
\frac{1}{4}g_{mn}\dd \Phi^m \dd\Phi^n &= G_{i\bar j} \dd \tau^i \dd \bar \tau^{\bar j}+\dd U^2+\frac{1}{4} e^{-4 U}(\dd \sigma + \tilde \zeta_I \dd \zeta^I-\zeta^I \dd \tilde \zeta_I)^2\\
&+\frac{1}{2}e^{-2U} \left[ \text{Im}\mathcal{N}_{IJ} \dd \zeta^I \dd \zeta^J+(\text{Im}\mathcal{N})^{-1}_{IJ}(\dd \tilde \zeta_I-\text{Re}\mathcal{N}_{IK}\dd \zeta^K)(\dd \tilde \zeta_J-\text{Re}\mathcal{N}_{JL}\dd \zeta^L) \right].
\eea

Note that the equations of motion coming from \eqref{eq:4daction} are invariant under the $SL(2,\mathbb{R})^{3\times}_U$ subgroup of the full U-duality group $E_{7(7)}$ mapping the STU model to itself. An element
\beq
S_1\otimes S_2 \otimes S_3 \in SL(2,\mathbb{R})^{3\times}_U,
\eeq
is parametrized as
\beq
S_i=
\left(
\begin{array}{cc}
a_i & b_i \\
c_i & d_i
\end{array}
\right),
\;\; a_i d_i-b_i c_i=1, \;\; i=1,2,3,
\eeq
and it acts on the 3 dimensional fields the following way. The scalars $\tau^i$ are transformed as
\bea
\tau^i &\mapsto \frac{a_i \tau^i+b_i}{c_i \tau^i +d_i},\\
\eea
while the electromagnetic potentials $\zeta^I$, $\tilde \zeta_I$ transform in the fundamental $(2,2,2)=8$ dimensional representation. Explicitely if one defines the three index tensor $\psi_{ijk}$ corresponding to the amplitudes of a 3 qubit state as
\beq
\label{eq:potentialUduality}
\left(
\begin{array}{cccc}
\psi_{000} & \psi_{001} & \psi_{010} & \psi_{011}\\
\psi_{100} & \psi_{101} & \psi_{110} & \psi_{111}\\ 
\end{array}
\right)=
\left(
\begin{array}{cccc}
\tilde \zeta_4 & \zeta^3 & \zeta^2 & -\tilde \zeta_1\\
\zeta^1 & -\tilde \zeta_2 & -\tilde \zeta_3 & -\zeta^4\\ 
\end{array}
\right)
\eeq
 the transformation rule is
 \beq
\psi_{ijk}\mapsto {(S_1)_i}^{i'} {(S_2)_j}^{j'} {(S_3)_k}^{k'}\psi_{i'j'k'}.
\eeq
In the following when we write $SL(2)$ we are referring ot $SL(2,\mathbb{R})$ unless explicitly otherwise stated.

We can describe the second term in the 3d Lagrangian \eqref{eq:3daction} as a sigma model with target space $SO(4,4)/SL(2)^{\times 4}$. The Lie algebra $\mathfrak{so}(4,4)$ has 28 generators. To describe this coset model we have to split this as $\mathfrak{so}(4,4)=\mathfrak{h}\oplus \mathfrak{m}$, where $\mathfrak{h}$ is an $\mathfrak{sl}_2^{\times 4}$ subalgebra and $\mathfrak{m}$ is its 16 dimensional fundamental representation which we denote as $(2,2,2,2)$. There are actually three ways to perform this split. We summarize bellow two of them which are relevant for our purposes.
\bea
\label{eq:splitsummary}
(2,2,2,2)  & \big\{  && H_\Lambda, && p_\Lambda=E_\Lambda+F_\Lambda && | && p^{Q_I}=E^{Q_I}-F^{Q_I}, && p^{P^I}=E^{P^I}-F^{P^I}  \\ \hline
\mathfrak{sl}_2^{\times 4} & \big\{  \undermat{(\mathfrak{sl}_2^{\times 4})_U}{&&  && k_\Lambda=E_\Lambda-F_\Lambda && }| \undermat{(2,2,2,2)_U}{ && k^{Q_I}=E^{Q_I}+F^{Q_I},  && k^{P^I}=E^{P^I}+F^{P^I}  } \\
& && && && && &&
\eea
Here, $H_\Lambda$, $E_\Lambda$, $F_\Lambda$, $\Lambda=0,...,3$ and $E^{Q_I}$, $E^{P^I}$, $F^{Q_I}$, $F^{P^I}$, $I=1,...,4$ are the 28 generators of $\mathfrak{so}(4,4)$. For an explicit definition of these generators and a detailed review of the splits, we refer to Appendix \ref{sec:liealg}. The subscript $U$ refers to the splitting suited to describe the action of the U-duality group $SL(2)^{\times 3}$. The extra $SL(2)$ factor in this split is the Ehlers symmetry. The subalgebras without the subscript answer the split suited to describe the 3d fields parametrizing the coset $SO(4,4)/SL(2)^{\times 4}$.

Let us represent an element of this coset by an $SO(4,4)$ matrix as\cite{bergshoeff2009generating,compere2}
\beq
\label{eq:cosetelement}
\mathcal{V}=e^{-U H_0}e^{-\frac{1}{2}\sum_i \log y_i H_i}e^{-\sum_i x_i E_i}e^{-\sum_I(\zeta^I E^{Q_I}+\tilde \zeta_I E^{P^I})}e^{-\frac{1}{2}\sigma E_0},
\eeq
Notice that there are no $F$-type generators in the above formula. This choice of gauge is called the Iwasawa gauge.  The next step is to project the Mauer-Cartan 1-form to the 16 dimensional space spanned by the first line of \eqref{eq:splitsummary} as
\beq
P_*=\frac{1}{2}(\dd \mathcal{V} \mathcal{V}^{-1}+(\dd \mathcal{V} \mathcal{V}^{-1})^\#).
\eeq
Here, $^\#$ is the anti-involution defineing the horizontal split of \eqref{eq:splitsummary} (see also e.q. \eqref{eq:sharpdef}). One finds that the target space metric \eqref{eq:cosetmetric} can be written as the right invariant metric on $SO(4,4)/SL(2)^{\times 4}$:
\beq
g_{mn}\dd \Phi^m \dd \Phi^n = \text{Tr}(P_*^2).
\eeq
The other way of writing this line element is to define the matrix
\beq
\mathcal{M}=\mathcal{V}^\# \mathcal{V}.
\eeq
Then we have
\beq
g_{mn}\dd \Phi^m \dd \Phi^n = \frac{1}{4}\text{Tr}((\mathcal{M}^{-1} \dd \mathcal{M})^2).
\eeq
We note that in contrary to $P_*$, in general $\mathcal{M}^{-1} \dd \mathcal{M}$ is not required to sit inside the 16 dimensional subspace which is spanned by the first line of \eqref{eq:splitsummary}. 

The 16 three dimensional fields can be extracted directly from $\mathcal{M}$ (see \cite{compere2} for details) so we may proceed describing the theory in terms of $\mathcal{M}$. A group element $h\in SO(4,4)$ acts naturaly on $\mathcal{V}$ as
\beq
\mathcal{V}\mapsto q\mathcal{V} h,
\eeq
where $q\in SL(2)^{\times 4}$ is a (possibly field dependent) compensator which puts $\mathcal{V}$ back to the Iwasawa gauge. We will see an example of this in sec. \ref{sec:cosetUduality} when we work out the action of the U-duality group on $\mathcal{V}$. The same action in terms of $\mathcal{M}$ reads as
\beq
\mathcal{M} \mapsto h^\# \mathcal{M} h.
\eeq
It is then manifest that the line element and hence the 3d Lagrangian is invariat under this action of $SO(4,4)$.

\section{The most general non-extremal black hole solution}
\label{sec:bhsolution}

Here we give a lightning review of some of the results and the solution generating technique presented in \cite{compere2}. We have seen that if $\mathcal{M}$ is a solution to the equations of motion then so is $h^\# \mathcal{M} h$. This fact can be used to generate new solutions from a known seed. This method is used by the authors of \cite{compere2} to find the most general non-extremal, rotating, asymptotically Taub-NUT black hole solution with 11 independent conserved charges. These are the mass, NUT charge, angular momentum and 8 independent dyonic charges. They chose the four dimensional Kerr-Taub-NUT metric as their seed with mass $m$, NUT charge $n$ and angular momentum $J=ma$ (here $a$ is a parameter). Then they chose the group element
\beq
\label{eq:chargeup}
h=e^{-\sum_I \gamma_I k^{P^I}}e^{-\sum_I \delta_I k^{Q_I}},
\eeq
to charge up the solution as
\beq
\mathcal{M}=h^\# \mathcal{M}_{KTN} h.
\eeq
We refer to \cite{compere2} for the definition of the seed matrix $\mathcal{M}_{KTN}$. Here, we merely need how the physical charges of the black hole are expressed with the 11 parameters $m,n,a,\delta_I,\gamma_I$. These charges are extracted as follows. Define the inverse radial coordinate $\rho=\frac{1}{r}$ and expand the fields around asymptotic infinity. For now, we assume the following expansion
\bea
\label{eq:fieldexpansion}
e^{2 U} & = 1 - 2M\rho + O (\rho^{2}) , && \zeta^I & = Q_I \rho + O(\rho^{2}) , && y_i & = \left( 1-\Sigma_i \rho + O(\rho^{2})\right) ,\\
\sigma &=-4 N\rho+(4 J \cos \theta+c)\rho^2+O(\rho^3) , && \tilde \zeta_I & = P^I\rho+ O(\rho^{2}) , && x_i & = \Xi_i \rho + O(\rho^{2}) ,
\eea
 which is the same as in \cite{compere2}. Here $M$ and $J$ are the ADM mass and angular momentum of the spacetime, $Q_I$ and $P^I$ are electric and magnetic charges associated to the original 4d $U(1)$ gauge fields and $\Xi_i$, $\Sigma_i$ denote 6 scalar charges. The constant $c$ is not playing any role in the following. The corresponding expansion for $\mathcal{M}$ is
\beq
 \label{eq:chargematrix1}
 \mathcal{M}=I+Q\rho + O(\rho^2),
 \eeq
 where the charge matrix $Q$ does not contain the angulat momentum $J$ which only enters in subleading order. Without $J$, these are all together 16 asymptotic charges, which can be expressed in terms of 10 seed \& charge parameters, $m$, $n$, $\delta_I$, $\gamma_I$. Therefore, the scalar charges are not independent of $M$, $N$ and the dyonic charges, but we will keep them explicit until section \ref{sec:scalcharg}. 
 
  We quote the formula for the mass and the NUT charge
\bea
M=\mu_1 m+\mu_1 n, && N=\nu_1 m+\nu_2 n,
\eea
where $\mu_1$, $\mu_2$, $\nu_1$, $\nu_2$ are functions only of $\delta_I$, $\gamma_I$ and are given explicitly as
\bea
\label{eq:mu}
\mu_1 & = 1 + \sum_I \bigg( \frac{s_{\delta I}^2 + s_{\gamma I}^2}{2} - s_{\delta I}^2 s_{\gamma I}^2 \bigg) + \frac{1}{2} \sum_{I, J} s_{\delta I}^2 s_{\gamma J}^2 , \\
 \mu_2 & = \sum_I s_{\delta I} c_{\delta I} \bigg( \frac{s_{\gamma I}}{c_{\gamma I}} c_{\gamma 1 2 3 4} - \frac{c_{\gamma I}}{s_{\gamma I}} s_{\gamma 1 2 3 4}\bigg) ,
\eea
\bea
\label{eq:nu}
\nu_1 & = \sum_I s_{\gamma I} c_{\gamma I} \bigg( \frac{c_{\delta I}}{s_{\delta I}} s_{\delta 1 2 3 4} - \frac{s_{\delta I}}{c_{\delta I}} c_{\delta 1 2 3 4} \bigg) , & \nu_2 = \iota - D,
\eea
with
\bea
\iota &= c_{\delta 1 2 3 4}c_{\gamma 1 2 3 4}+s_{\delta 1 2 3 4} s_{\gamma 1 2 3 4}+ \sum_{I < J} c_{\delta 1 2 3 4} \frac{s_{\delta I J}}{c_{\delta I J}} \frac{c_{\gamma I J}}{s_{\gamma I J}} s_{\gamma 1 2 3 4} , \\
D &= c_{\delta 1 2 3 4}s_{\gamma 1 2 3 4}+s_{\delta 1 2 3 4}c_{\gamma 1 2 3 4} + \sum_{I < J} c_{\delta 1 2 3 4} \frac{s_{\delta I J}}{c_{\delta I J}} \frac{s_{\gamma I J}}{c_{\gamma I J}} c_{\gamma 1 2 3 4}.
\eea
Here the notation is resolved as follows: $c_{\delta_I}=\cosh \delta_I$, $s_{\delta_I}=\sinh \delta_I$ and the same for $\gamma_I$. Multiple indices denote that one should take the product of the hyperbolic functions e.g. $c_{\delta IJ}=\cosh \delta_I \cosh \delta_J$. The rest of the charges can be obtained through the formulae
\bea
\label{eq:deltaderiv1}
\frac{\partial 2 M}{\partial \delta_I} &=Q_I, && \frac{\partial 2 N}{\partial \delta_I} &=-P^I,
\eea
\beq
\label{eq:deltaderivQ}
\frac{\partial Q_I}{\partial \delta_J} = \left(\small{
\begin{array}{cccc}
2M-\Sigma_1+\Sigma_2+\Sigma_3 & & & \\
 & 2M+\Sigma_1-\Sigma_2+\Sigma_3 & & \\
 & & 2M+\Sigma_1+\Sigma_2-\Sigma_3 & \\
 & & & 2M-\Sigma_1-\Sigma_2-\Sigma_3
\end{array}}
\right),
\eeq
\beq
\frac{\partial P^I}{\partial \delta_J} = \left(
\begin{array}{cccc}
-2 N & \Xi_3 & \Xi_2 & -\Xi_1 \\
 \Xi_3 & -2 N & \Xi_1 & -\Xi_2\\
 \Xi_2 & -\Xi_1 & -2 N & -\Xi_3 \\
 -\Xi_1 & -\Xi_2 & -\Xi_3 & -2 N
\end{array}
\right), 
\eeq
\bea
\label{eq:deltaderiv2}
\frac{\partial \Sigma_i}{\partial \delta_J} &= \left(
\begin{array}{cccc}
-Q_1 & Q_2 & Q_3 & -Q_4 \\
 Q_1 & -Q_2 & Q_3 & -Q_4\\
 Q_1 & Q_2 & -Q_3 & -Q_4 \\
\end{array}
\right), &&
\frac{\partial \Xi_i}{\partial \delta_J} &= \left(
\begin{array}{cccc}
-P^4  & P^3 & P^2 & -P^1 \\
 P^3 & -P^4 & P^1 & -P^2\\
 P^2 & P^1 & -P^4 & -P^3 \\
\end{array}
\right).
\eea
Note that these identities are simple consequences of the fact that we have defined the charging up element $h$ with the $\delta_I$ parameters being on the right. Therefore, for the charge matrix $Q=h^\# Q_{KTN} h$ one has
\beq
\frac{\partial Q}{\partial \delta_I}=[k^{Q_I},Q],
\eeq
and hence taking the $\delta_I$ derivative of a component of $Q$ in some basis just amounts to multiplying with the adjoint representation of $k^{Q_I}$ in the same basis. We have used here that $h$ is generated by the subalgebra in the second line of \eqref{eq:splitsummary} and hence we have $h^\#=h^{-1}$.

After reconstructing the 4d solution the Bekenstein-Hawking entropy of the black hole can be calculated. It reads as\cite{compere2}
\beq
S=2\pi ( \sqrt{\Delta+F}+\sqrt{-J^2+F}),
\eeq
where $\Delta$ is the quartic invariant formed from the dyonic charges, and $F$ is expressed with the seed and charge parameters as
\beq
F=(m^2+n^2)(m \nu_2- n\nu_1)^2.
\eeq
To obtain asymptotically flat black holes one can cancel the NUT charge by setting $n=-m\frac{\nu_1}{\nu_2}$.

\section{The action of the U-duality group}
\label{sec:actionofU}

In this section we describe in detail the transformation properties of the asymptotic charges of the black hole under U-duality. This allows us to construct polynomial invariants of these in the next section.

\subsection{Action on fields}
\label{sec:cosetUduality}

 Recall that the coset element is parametrized in Iwasawa gauge as
\beq
\mathcal{V}=e^{-U H_0}e^{-\frac{1}{2}\sum_i \log y_i H_i}e^{-\sum_i x_i E_i}e^{-\sum_I(\zeta^I E^{Q_I}+\tilde \zeta_I E^{P^I})}e^{-\frac{1}{2}\sigma E_0}.
\eeq
Notice that the gauge is such that the scalar fields are in an "upper triangular" form in the sence that the $F$ type generators are not present. We claim that the U-duality group is generated by the generators $H_i$, $E_i$, $F_i$, $i=1,2,3$ of \eqref{eq:splitsummary} and the action is a simple right action on $\mathcal{V}$ followed by the left action of a local compensator $q\in SO(2)^{\times3} \subset SL(2)^{\times 3}_U$ restoring the Iwasawa gauge:
\beq
\mathcal{V} \mapsto q\mathcal{V}g, \;\; g\in SL(2)^{\times 3}_U.
\eeq
Indeed, we may write this as
\beq
 q\mathcal{V}g=e^{-U H_0}\left(q e^{-\frac{1}{2}\sum_i \log y_i H_i}e^{-\sum_i x_i E_i}g\right) \left( g^{-1}e^{-\sum_I(\zeta^I E^{Q_I}+\tilde \zeta_I E^{P^I})}g\right) e^{-\frac{1}{2}\sigma E_0},
\eeq
as the generators $E_0$ and $H_0$ commute with $SL(2)^{\times 3}_U$. We see that the part with the potentials transform simply with the adjoint action. It is convenient to describe the 16 dimensional representation $(2,2,2,2)_U$ and its element $\sum_I(\zeta^I E^{Q_I}+\tilde \zeta_I E^{P^I})$ with the amplitudes $\psi_{ijkl}$ of a four qubit state through \eqref{eq:Udual4qubit}. Since there are no $F^{P^I}$ and $F^{Q_I}$ generators present this four qubit state has $\psi_{1jkl}\equiv 0$ and therefore it is actually a three qubit state. We can think of $\psi_{1jkl}\equiv 0$ as the gauge fixing condition. Then, under the U-duality transformation
\beq
\sum_I(\zeta^I E^{Q_I}+\tilde \zeta_I E^{P^I}) \mapsto g^{-1}\sum_I(\zeta^I E^{Q_I}+\tilde \zeta_I E^{P^I})g, \;\;\; g\in SL(2)^{\times 3}_U,
\eeq
this associated state transform as
\beq
\label{eq:chargeSLOCC}
\psi_{ijkl}\mapsto {(S_1)_j}^{j'} {(S_2)_k}^{k'} {(S_3)_l}^{l'}\psi_{ij'k'l'}, \;\;  S_1\otimes S_2\otimes S_3 \in SL(2)^{\times 3}_U,
\eeq
where, according to \eqref{eq:Udual4qubit}, $S_1\otimes S_2\otimes S_3$ is the $SL(2)^{\times 3}_U$ element associated to $g^{-1}$. We see that the gauge condition $\psi_{1jkl}\equiv 0$ on the coset element does not change. Therefore, this gives the required action of the U-duality group on the potentials given in \eqref{eq:potentialUduality} and no compensator is needed. We only need to worry about the gauge condition on the scalars. Multiplying the scalar term from the right with $g$ spoils the gauge we chose and hence we need a local compensator. Using the fact that each scalar parametrizes the coset space $SL(2)/SO(2)$ we expect the local compensator to be from $SO(2)^{\times 3}$\cite{carbone} and therefore to have the form
\beq
q=e^{\sum_{i=1}^3 \alpha_i k_i}=e^{\sum_{i=1}^3 \alpha_i(E_i-F_i)}.
\eeq
Now it is easy to verify that if we let $g$ to be the $SL(2)^{\times 3}_U$ element corresponding to 
\beq
\label{SL23element}
\left(
\begin{array}{cc}
 d_1 & -b_1 \\
 -c_1 & a_1 \\
\end{array}
\right) \otimes
\left(
\begin{array}{cc}
 d_2 & -b_2 \\
 -c_2 & a_2 \\
\end{array}
\right)
\otimes
\left(
\begin{array}{cc}
 d_3 & -b_3 \\
 -c_3 & a_3 \\
\end{array}
\right),
\eeq
in the standard representation, then in order to restore the Iwasawa gauge the $\alpha_i$ of the compensator have to be chosen as
\beq
\label{eq:Ucompensator}
\tan \alpha_i=\frac{c_i y_i}{c_i x_i+a_i}.
\eeq
Then, we have
\beq
q e^{-\frac{1}{2}\sum_i \log y_i H_i}e^{-\sum_i x_i E_i}g = e^{-\frac{1}{2}\sum_i \log y_i' H_i}e^{-\sum_i x_i' E_i},
\eeq
with the primed scalars being
\bea
x_i'=\frac{(d_i+c_i x_i)(b_i+a_i x_i)+a_i c_i(x_i^2+y_i^2)}{(d_i+c_i x_i)^2+c_i^2 y_i^2}, &&
y_i'=\frac{y_i}{(d_i+c_i x_i)^2+c_i^2 y_i^2},
\eea
which just corresponds to the usual action of the U-duality group on the scalars
\beq
\tau_i'=\frac{a_i \tau_i+b_i}{c_i \tau_i+d_i},
\eeq
with $\tau_i=x_i+i y_i$. 

\subsection{Action on asymptotic charges}
\label{sec:Uoncharges}

Now recall that the asymptotic values of the fields are conveniently encoded in a charge matrix $Q$ (see \eqref{eq:chargematrix1}) defined from the series expasion of $\mathcal{M}=\mathcal{V}^\# \mathcal{V}$ around asymptotic infinity. Now we need to be slightly more general than in \cite{compere2} and cover the moduli space by letting $\mathcal{M}(\rho=0)\neq I$. To define this generalized $Q$ first expand $\mathcal{M}$ around asymptotic infinity 
\beq
\label{eq:cosetMexp}
\mathcal{M}=\mathcal{M}^{(0)}+\mathcal{M}^{(1)}\rho+ O(\rho^2).
\eeq
Then note that $\mathcal{M}=\mathcal{V}^\# \mathcal{V}$ is a proper element of $SO(4,4)$ for every value of the coordinates. It follows that $\mathcal{M}^{(0)}$ and $(\mathcal{M}^{(0)})^{-1}\mathcal{M}$ are all good $SO(4,4)$ elements. Expanding this latter yields
\beq
(\mathcal{M}^{(0)})^{-1}\mathcal{M}=I+Q\rho+ O(\rho^2),
\eeq
where we have defined the "dressed" charge matrix\cite{cvetic}
\beq
\label{eq:chargematrix}
Q=(\mathcal{M}^{(0)})^{-1}\mathcal{M}^{(1)}.
\eeq
As this is an expansion of an element of $SO(4,4)$ around the identity, $Q$ is an element of the Lie algebra $\mathfrak{so}(4,4)$. This is not true for $\mathcal{M}^{(1)}$ alone. Now we slightly generalize the expansion \eqref{eq:fieldexpansion} of the fields around asymptotic infinity by letting
\bea
\label{eq:generalasympt}
 y_i & = Y_i\left( 1-\Sigma_i \rho + O(\rho^{2})\right)  , && x_i & = X_i+\Xi_i \rho + O(\rho^{2}) ,
\eea
with the rest being the same as in \eqref{eq:fieldexpansion}, but now with the possibility of having arbitrary values for the scalar fields at infinity. We give this general $Q$ matrix in section \ref{app:generalscalars}. and for now, go back to use $X_i=0$ and $Y_i=1$ as in \cite{compere2}. However, we stress that $X_i$ and $Y_i$ do transform under U-duality. One should not worry about this too much as everything we do in the following relies only on the fact that $Q$ is an element of $\mathfrak{so}(4,4)$ which is independent of this choice and hence straightforwardly generalize to arbitrary $X_i$ and $Y_i$. Setting $X_i=0$ and $Y_i=1$ results in the simple form of $Q$
\beq
\label{eq:Qexpand}
Q=2M H_0 +\sum_i \Sigma_i H_i +2N p_0-\sum_i \Xi_i p_i-\sum_I(Q_I p^{Q_I}+P^Ip^{P^I}).
\eeq
We see that in this case $Q$ lives in the 16 dimensional subspace $(2,2,2,2)$ spanned by the first line of \eqref{eq:splitsummary}. 

From the previous subsection we conclude that $\mathcal{M}$ transforms under U-duality as $\mathcal{M} \mapsto g^\# \mathcal{M} g$, and hence the charge matrix simply transforms with the adjoint action
\beq
Q\mapsto g^{-1}Qg, \;\;\; g\in SL(2)^{\times 3}_U.
\eeq
Now from \eqref{eq:Qexpand} we easily see that under the decomposition \eqref{eq:splitsummary} of $\mathfrak{so}(4,4)$ suitable for U-duality, $Q$ has components both in ${\mathfrak{sl}_2^{\times 4}}_U$ and $(2,2,2,2)_U$. As the splitting ensures that these components do not mix under the adjoint action of ${\mathfrak{sl}_2^{\times 4}}_U$ we may consider these parts separately
\beq
Q=Q_{-}+Q_{+},
\eeq
where
\bea
Q_{-}&=2M H_0 +\sum_i \Sigma_i H_i +2N p_0-\sum_i \Xi_i p_i \in {\mathfrak{sl}_2^{\times 4}}_U,\\
Q_{+}&=-\sum_I(Q_I p^{Q_I}+P^Ip^{P^I}) \in (2,2,2,2)_U.
\eea
Let us first consider $Q_{-}$. Clearly, $M$ and $N$ are invariant under the adjoint action of $SL(2)^{\times 3}_U$ as expected. The six scalar charges $\Sigma_i$, $\Xi_i$ parametrize an element ${\mathfrak{sl}_2^{\times 3}}_U$ and hence they transform in the \textit{adjoint representation} of $SL(2)^{\times 3}_U$. We define the matrices
\beq
\label{eq:Ris}
R_i=\left(
\begin{array}{cc}
 \Sigma_i & -\Xi_i \\
-\Xi_i  & -\Sigma_i \\
\end{array}
\right),
\eeq
transforming under U-duality as
\beq
R_i \mapsto \left(
\begin{array}{cc}
 a_i & b_i\\
c_i & d_i \\
\end{array}
\right)
R_i
\left(
\begin{array}{cc}
 a_i & b_i\\
c_i & d_i \\
\end{array}
\right)^{-1},
\eeq
with $\left(
\begin{array}{cc}
 a_i & b_i\\
c_i & d_i \\
\end{array}
\right)\in SL(2)$. Note that the symmetricity of $R_i$ is spoiled by this transformation but this is simply a consequence of fixing the asymptotic values of the scalars which, again, are not invariant under U-duality (see section \ref{app:generalscalars}. for the general form of $R$).

The element $Q_{+}$ transforms as a four qubit state $\psi_{ijkl}$ under the full $SL(2)^{\times 4}_U$ and it decomposes into a pair of three qubit states $(\psi_1)_{jkl}\equiv \psi_{0jkl}$ and $(\psi_2)_{jkl}\equiv \psi_{1jkl}$ when just the U-duality group $SL(2)^{\times 3}_U$ is used. The explicit amplitudes can be read off using \eqref{eq:Udual4qubit} and are given as
\beq
\label{eq:pairofthreequbits}
\left(
\begin{array}{cccc}
\psi_{0000} & \psi_{0001} & \psi_{0010} & \psi_{0011}\\
\psi_{0100} & \psi_{0101} & \psi_{0110} & \psi_{0111}\\ \hline
\psi_{1000} & \psi_{1001} & \psi_{1010} & \psi_{1011}\\
\psi_{1100} & \psi_{1101} & \psi_{1110} & \psi_{1111}
\end{array}
\right)=
\left(
\begin{array}{cccc}
-P^4 & -Q_3 & -Q_2 & P^1\\
-Q_1 & P^2 & P^3 & Q_4\\ \hline
-Q_4 & P^3 & P^2 & Q_1\\
P^1 & Q_2 & Q_3 & -P^4
\end{array}
\right).
\eeq
Note that this pair is related through
\beq
\label{eq:relatedqubits}
|\psi_2\rangle = \left(
\begin{array}{cc}
0 & -1\\
1 & 0
\end{array}
\right)
\otimes
\left(
\begin{array}{cc}
0 & -1\\
1 & 0
\end{array}
\right)
\otimes
\left(
\begin{array}{cc}
0 & -1\\
1 & 0
\end{array}
\right)|\psi_1 \rangle.
\eeq
For the corresponding pair of three qubit states dressed with non-trivial scalar asymptotics, see again section \ref{app:generalscalars}. 

The index corresponding to the first qubit transforms as a doublet under the extra Ehlers $SL(2)$. Note that the scalar charges are singlets under the Ehlers symmetry: the adjoint $28$ of $SO(4,4)$, where $Q$ lives in, decomposes under the maximal subgroup $SL(2)^{\times 4}_U=SL(2)_{Ehlers}\times (SL(2)^{\times 3}_U)$ as $28=(3,1)\oplus (1,9) \oplus (2,8)$. 

\subsection{The algebra of covariants}

We have seen that the asymptotic charges transform under U-duality as a (not general) vector in $9\oplus 8\oplus 8$, where $9$ refers to the adjoint representation, while $8$ is the fundamental corresponding to a three qubit state. If we dress up the charge matrix with the asymptotic values of the scalars, then $Q$ actually fills out the representation $9\oplus 8\oplus 8$. In order to be able to write up invariants we first need to construct covariants with indices transforming the same way. Luckily, there exists a construction, called the moment map, which allows one to associate an element transforming in $9$ to a \textit{pair} of vectors in 8. Unfortunately, this construction will result in an unnecessaryly large covariant algebra. We can significantly reduce this by incorporating "triality" symmetry of the STU model: the symmetry under permutation of the three $SL(2)$ factors. This leads us to consider the embedding ${\mathfrak{sl}_2^{\times 3}}_U \subset \mathfrak{sl}_6$ and to construct the moment map in the $SL(6)$ covariant language of three fermions with six single particle states\cite{levvran1,levsar1,levsar2}. 

Consider fermionic creation an annihilation operators $p^a$ and $n_a$, $a=1,...,6$ satisfying
\bea
\lbrace p^a,n_b\rbrace={\delta^a}_b, && \lbrace p^a,p^b\rbrace=0, && \lbrace n_a,n_b\rbrace=0.
\eea
An unnormalized three fermion state can be written as
\beq
|P\rangle = \frac{1}{3!}P_{abc}p^a p^b p^c|0\rangle \in \wedge^3 (\mathbb{C}^6),
\eeq
with the antisymmetric tensor $P_{abc}$ having 20 independent components. The so-called SLOCC group of this system is $SL(6,\mathbb{C})$ acting locally on the amplitudes as
\beq
\label{eq:fermionSLOCC}
P_{abc}\mapsto {S_a}^{a'}{S_b}^{b'}{S_c}^{c'}P_{a'b'c'}, \;\;\; S\in SL(6,\mathbb{C}).
\eeq
The moment map associates an $\mathfrak{sl}(6,\mathbb{C})$ element to $|P\rangle$. This element reads as
\beq
{(K_P)^a}_b=\frac{1}{2!3!}\epsilon^{ac_1c_2c_3c_4c_5}P_{bc_1c_2}P_{c_3c_4c_5}.
\eeq 
It is clear that if we transform the state as in \eqref{eq:fermionSLOCC} this covariant transforms as
\beq
K_P \mapsto (S^T)^{-1}K_P S^T,
\eeq
hence the powers of its trace are continous invariants. It turns out that the action of $SL(6,\mathbb{C})$ on $20$ admits a single independent continous invariant, quartic in the amplitudes, given by
\beq
\mathcal{D}(P)=\frac{1}{6}\text{Tr}K_P^2.
\eeq
Note that this quantity is a measure of tripartite entanglement for the fermions\cite{levvran1,levsar3}. The situation is different if we have two states $|P\rangle$ and $|Q\rangle$ at our disposal. In this case we can define the following covariants
\bea
\label{eq:pairoffermions}
{(K_P)^a}_b&=\frac{1}{2!3!}\epsilon^{ac_1c_2c_3c_4c_5}P_{bc_1c_2}P_{c_3c_4c_5},\\
{(K_Q)^a}_b&=\frac{1}{2!3!}\epsilon^{ac_1c_2c_3c_4c_5}Q_{bc_1c_2}Q_{c_3c_4c_5},\\
{(K_{PQ})^a}_b&=\frac{1}{2!3!}\epsilon^{ac_1c_2c_3c_4c_5}P_{bc_1c_2}Q_{c_3c_4c_5},\\
{(K_{QP})^a}_b&=\frac{1}{2!3!}\epsilon^{ac_1c_2c_3c_4c_5}Q_{bc_1c_2}P_{c_3c_4c_5}.
\eea
Traces of products of these define invariants. In particular there is now a non-zero invariant bilinear product
\beq
\label{eq:fermioninnerprod}
(P,Q)=\frac{1}{3}\text{Tr}K_{PQ}=-\frac{1}{3}\text{Tr}K_{QP}=\frac{1}{3!3!}\epsilon^{c_1c_2c_3c_4c_5c_6}P_{c_1c_2c_3}Q_{c_4c_5c_6},
\eeq
which also shows that the covariants $K_{PQ}$ and $K_{QP}$ are now \textit{not} elements of the Lie algebra $\mathfrak{sl}_6$.

Now let us describe how to embed three qubit states
\beq
|\psi\rangle=\sum_{i,j,k=0}^1 \psi_{ijk}|ijk\rangle,
\eeq
into our fermionic vector space. The standard thing to do is the following:
\beq
\label{eq:fermionsfromqubits}
|\psi\rangle \mapsto |P_\psi\rangle =\sum_{i,j,k=0}^1 \psi_{ijk}p^{i+1}p^{j+3}p^{k+5}|0\rangle.
\eeq
Then, it is easy to see that a three qubit SLOCC transformation
\beq
\psi_{ijk}\mapsto {(S_1)_i}^{i'} {(S_2)_j}^{j'} {(S_3)_k}^{k'}\psi_{ij'k'}, \;\;  S_1\otimes S_2\otimes S_3 \in SL(2)^{\times 3}_U,
\eeq
can be implemented in the language of of three fermions \eqref{eq:fermionSLOCC} by choosing
\beq
S=\left(
\begin{array}{ccc}
S_1 & & \\
 & S_2 & \\
 & & S_3
\end{array}
\right)\in SL(6,\mathbb{C}).
\eeq

Finally, we can associate covariants transforming in the adjoint of $SL(2)^{\times 3}_U$ to the pair of three qubit states given in \eqref{eq:pairofthreequbits} using \eqref{eq:pairoffermions} and \eqref{eq:fermionsfromqubits} as
\bea
\label{eq:fundcovariants}
{(K_{11})^a}_b&=\frac{1}{2!3!}\epsilon^{ac_1c_2c_3c_4c_5}(P_{\psi_1})_{bc_1c_2}(P_{\psi_1})_{c_3c_4c_5},\\
{(K_{22})^a}_b&=\frac{1}{2!3!}\epsilon^{ac_1c_2c_3c_4c_5}(P_{\psi_2})_{bc_1c_2}(P_{\psi_2})_{c_3c_4c_5},\\
{(K_{12})^a}_b&=\frac{1}{2!3!}\epsilon^{ac_1c_2c_3c_4c_5}(P_{\psi_1})_{bc_1c_2}(P_{\psi_2})_{c_3c_4c_5},\\
{(K_{21})^a}_b&=\frac{1}{2!3!}\epsilon^{ac_1c_2c_3c_4c_5}(P_{\psi_2})_{bc_1c_2}(P_{\psi_1})_{c_3c_4c_5}.
\eea
For an explicit form of these matrices see appendix \ref{app:stumatrices}. Notice the crucial fact that these four matrices transform exactly the same way under $SL(2)^{\times 3}_U$ as the matrix
\beq
\label{eq:Rmatrix}
R =
\left(
\begin{array}{ccc}
R_1^T & & \\
 & R_2^T & \\
  & & R_3^T
\end{array}
\right)
=\left(
\begin{array}{cccccc}
 \Sigma_1 & -\Xi_1 & 0 & 0 & 0 & 0 \\
 -\Xi_1 & -\Sigma_1 & 0 & 0 & 0 & 0 \\
 0 & 0 & \Sigma_2 & -\Xi_2 & 0 & 0 \\
 0 & 0 & -\Xi_2 & -\Sigma_2 & 0 & 0 \\
 0 & 0 & 0 & 0 & \Sigma_3 & -\Xi_3 \\
 0 & 0 & 0 & 0 & -\Xi_3 & -\Sigma_3 \\
\end{array}
\right),
\eeq
formed by the scalar charges (see e.q. \eqref{eq:Ris}).

The four matrices of \eqref{eq:fundcovariants} can be grouped into a $2\times 2$ block matrix $K_{ab}$ transforming under the Ehlers $SL(2)$ as $2\times 2=1\oplus 3$ in its $ab$ indices. We note that the singlet part satisfies the relation
\beq
K_{21}-K_{12}=\bigg( \sum_I(Q_I^2+(P^I)^2) \bigg) I,
\eeq
and hence only three of the $K_{ab}$s give independent covariants. 

\section{The $F$-invariant}
\label{sec:Finv}

\subsection{Construction of the invariant}

Let us begin by listing the independent primitive invariants of homogeneous degree less than or equal to four in the asymptotic charges, that can be formed by our covariants.
\begin{itemize}
\item Degree 1 invariants:
\bea
M, && N.
\eea
\item Degree 2 invariants:
\bea
\label{eq:priminv1}
\text{Tr}(K_{12}), && \text{Tr}(R^2).
\eea
\item Degree 3 invariants:
\bea
\text{Tr}(K_{12} R), && \text{Tr}(K_{11} R).
\eea
\item Degree 4 invariants:
\bea
\label{eq:priminv2}
\text{Tr}K_{11}^2, && \text{Tr}K_{12}^2, && \text{Tr}(K_{11}K_{22}), && \text{Tr}(R^4).
\eea
\end{itemize}
To reduce the set of independent invariants we have used the identities
\bea
\text{Tr}(K_{11}R)=-\text{Tr}(K_{22}R), && \text{Tr}(K_{11}^2)=\text{Tr}(K_{22}^2), \\
 \text{Tr}(K_{11}R^2)=\text{Tr}(K_{22}R^2)=0, &&
\text{Tr}(K_{12} K_{11})=-\text{Tr}(K_{12} K_{22}), \\
 \text{Tr}(K_{12}R^2)=\frac{1}{6}(\text{Tr}K_{12})(\text{Tr}R^2), &&
3\text{Tr}(K_{11}K_{22})+(\text{Tr}K_{12})^2-3\text{Tr}(K_{12}^2)=0.
\eea

Note that we can write some well-known U-duality invariants in this language. First of all, Cayley's hyperdeterminant
\beq
\label{eq:cayley}
\Delta= \frac{1}{16} \bigg( 4 (Q_1 Q_2 Q_3 Q_4 + P^1 P^2 P^3 P^4) + 2 \sum_{J < K} Q_J Q_K P^J P^K - \sum_J (Q_J)^2 (P^J)^2 \bigg) 
\eeq
is just given as
\beq
\Delta=-\frac{1}{96}\text{Tr}K_{11}^2=-\frac{1}{96}\text{Tr}K_{22}^2.
\eeq
The asymptotic value of the quadratic symplectic invariant $I_2$ is
\beq
I_2^\infty =\frac{1}{4}\sum_I(Q_I^2+(P^I)^2)=-\frac{1}{12}\text{Tr}K_{12}.
\eeq
The quadratic invariant
\beq
S_2^\infty = \frac{1}{4}G_{i\bar j}\partial_r \tau^i \partial_r \bar \tau^{\bar i}|_{r\rightarrow \infty}=\frac{1}{4}\sum_i(\Xi_i^2+\Sigma_i^2),
\eeq
can be expressed as
\beq
S_2^\infty=\frac{1}{8}\text{Tr}(R^2).
\eeq

Now recall the formula for the $F$ invariant in terms of the charge-up parameters $\delta_I$, $\gamma_I$ and seed variables $m$ and $n$:
\beq
F=(m^2+n^2)(m \nu_2- n\nu_1)^2,
\eeq
where $\nu_1$ and $\nu_2$ are the functions given in e.q. \eqref{eq:nu} and they do not scale with the charges. We see that $F$ is of homogeneous degree 4 in the charges and we expect it to be U-duality invariant.  From the 10 invariants that we have identified at the begining of this section we can form 22 monomials of homogeneous degree four. We can form a linear combination of these, equate it to $F$ and try to solve for the coefficients. The simplest way to do this is to generate various random sets of parameters $m$, $n$, $\delta_I$ and $\gamma_I$ and try to solve the resulting numerical, linear equations simultenaously. If this works for a considerably higher number of equations than the number of variables, which is 22, we can probably trust our coefficients. We did this procedure for 600 equations and we have found a single solution. The obtained numerical coefficients have been rationalized and the result was tested analyticaly with a computer algebra system. The result is
\bea
\label{eq:Finv1}
F&=M^4+M^2 N^2+\frac{ M^2}{12}\text{Tr}  K_{12}-\frac{M }{24}\text{Tr} ( K_{12}R)+\frac{N^2}{24}\text{Tr}(R^2)\\
&-\frac{N}{24}\text{Tr}(K_{11} R)+\frac{1}{192}\left(\text{Tr}(K_{11}^2)-\text{Tr}(K_{11}K_{22})-\frac{1}{2}(\text{Tr}R^2)^2+\text{Tr}(R^4)\right).
\eea
In the asymptotically flat case one sets $N=0$ (or $n=-m\frac{\nu_1}{\nu_2}$). In this case the $F$ invariant reads as
\bea
F = m^4 \frac{(\nu_1^2+\nu_2^2)^3}{\nu_2^4} &=M^4-M^2 I_2^\infty-\frac{M }{24}\text{Tr} ( K_{12}R)-\frac{1}{2}\Delta \\
 & -\frac{1}{192}\text{Tr}(K_{11}K_{22})-\frac{1}{6}(S_2^\infty)^2 +\frac{1}{192}\text{Tr}(R^4),
\eea
where we have reintroduced the familiar U-duality invariants where it is possible. We stress that for general scalar asymptotics one should use the $R$ and $K_{ab}$ matrices as given in section \ref{app:generalscalars}. It is useful to write the $F$ invariant without an explicit reference to the auxilary 6 dimensional representation that we have introduced. We may employ the invariant bilinear product of \eqref{eq:fermioninnerprod} to write
\bea
\text{Tr}  K_{12} &=3(P_{\psi_1},P_{\psi_2}),\\
\text{Tr} ( K_{12}R) &=-(P_{\psi_1},R_* P_{\psi_2}),\\
\text{Tr}(K_{11}K_{22})&=-(P_{\psi_1},(K_{22})_* P_{\psi_1}),\\
\text{Tr}(K_{11}^2)&=-(P_{\psi_1},(K_{11})_* P_{\psi_1}),
\eea
where we have defined the action of a Lie algebra element $t\in \mathfrak{sl}(6)$ on $P\in \wedge^3 \mathbb{C}$ as
\beq
(t_*P)_{abc}={t^d}_a P_{dbc}+{t^d}_b P_{adc}+{t^d}_c P_{abd}.
\eeq
Also, let us define the $8\times 8$ matrix $\hat R$ corresponding to $R$ in the fundamental representation of the U-duality group:
\beq
\label{eq:Rfundamental}
\hat R=R_1^T \otimes I \otimes I+I\otimes R_2^T \otimes I+I\otimes I\otimes R_3^T,
\eeq
see \eqref{eq:Ris} for the definition of $R_i$s. Then we have
\beq
\text{Tr}R^4-\frac{1}{2}(\text{Tr}R^2)^2 =-\frac{1}{8}\left( \text{Tr}\hat R^4-\frac{1}{8}(\text{Tr}\hat R^2)^2 \right).
\eeq
We may then rewrite the $F$ invariant as
\bea
\label{eq:Finv2}
F&=M^4+M^2 N^2+\frac{ M^2}{4}(P_{\psi_1},P_{\psi_2})+\frac{M }{24}(P_{\psi_1},R_* P_{\psi_2})+\frac{N^2}{96}\text{Tr}(\tilde R^2)\\
&+\frac{N}{24}(P_{\psi_1},R_* P_{\psi_1})-\frac{1}{192}\big(P_{\psi_1},(K_{11}-K_{22})_* P_{\psi_1}\big)\\
&+\frac{1}{1536}\left(\frac{1}{8}(\text{Tr}\hat R^2)^2-\text{Tr}(\hat R^4)\right),
\eea
which will be well suited for generalization to the $E_7$ invariant case.

As a final remark, we note that a single centered STU back hole parametrized by six moduli and eight dyonic charges is expected to have five independent U-duality invariants\cite{andrianopoli2014extremal}. The reason that we have more than this in \eqref{eq:priminv1}-\eqref{eq:priminv2} is that we treat the scalar charges as independent variables. This allowed us to turn F into a polynomial invariant.

\subsection{General scalar asymptotics}
\label{app:generalscalars}

Here we consider explicitly the computation of $F$ for general asymptotic values $X_i$ and $Y_i$ of the moduli. We expand the "dressed" charge matrix defined in \eqref{eq:chargematrix} using \eqref{eq:generalasympt} as
\beq
Q=q_{H_\Lambda}H_\Lambda+ q_{E_\Lambda}E_\Lambda+ q_{F_\Lambda}F_\Lambda +q_{E^{Q_I}}E^{Q_I}+q_{E^{P^I}}E^{P^I}+q_{F^{Q_I}}F^{Q_I}+q_{F^{P^I}}F^{P^I},
\eeq
Note that in general this matrix does not live in the 16 dimensional four qubit subspace corresponding to the first line of \eqref{eq:splitsummary}. The part in ${\mathfrak{sl}_2^{\times 4}}_U$ reads as
\beq
Q_-=2 M H_0+2 N p_0 + \sum_{i=1}^3 \left[ \left( \Sigma_i-\frac{\Xi_i X_i}{Y_i^2}\right)H_i
+  \left( -\Xi_i-2 \Sigma_i X_i+\frac{\Xi_i X_i^2}{Y_i^2} \right)E_i -\frac{\Xi_i}{Y_i^2} F_i\right],
\eeq
and hence the $R_i$ matrices of \eqref{eq:Ris} obtain the following dressing
\beq
\label{eq:dressedRis}
R_i=\left(
\begin{array}{cc}
 \Sigma_i-\frac{\Xi_i X_i}{Y_i^2} & -\Xi_i-2 \Sigma_i X_i+\frac{\Xi_i X_i^2}{Y_i^2}\\
-\frac{\Xi_i}{Y_i^2}  & -\Sigma_i +\frac{\Xi_i X_i}{Y_i^2} \\
\end{array}
\right).
\eeq
The four qubit state in $(2,2,2,2)_U$ is
\beq
Q_+=q_{E^{Q_I}}E^{Q_I}+q_{E^{P^I}}E^{P^I}+q_{F^{Q_I}}F^{Q_I}+q_{F^{P^I}}F^{P^I}.
\eeq
The amplitudes corresponding to the first three qubit state (see \eqref{eq:Udual4qubit}) are unchanged
\beq
\label{eq:dressedfirstqubit}
\left(
\begin{array}{cccc}
\psi_{0000} & \psi_{0001} & \psi_{0010} & \psi_{0011}\\
\psi_{0100} & \psi_{0101} & \psi_{0110} & \psi_{0111}\\ 
\end{array}
\right)\equiv
\left(
\begin{array}{cccc}
q_{E^{P^4}} & q_{E^{Q_3}} & q_{E^{Q_2}} & -q_{E^{P^1}}\\
 q_{E^{Q_1}} & -q_{E^{P^2}} & -q_{E^{P^3}} & -q_{E^{Q_4}} \\ 
\end{array}
\right)
=
\left(
\begin{array}{cccc}
-P^4 & -Q_3 & -Q_2 & P^1\\
-Q_1 & P^2 & P^3 & Q_4
\end{array}
\right).
\eeq
On the other hand the second three qubit state 
\beq
\left(
\begin{array}{cccc}
\psi_{1000} & \psi_{1001} & \psi_{1010} & \psi_{1011}\\
\psi_{1100} & \psi_{1101} & \psi_{1110} & \psi_{1111}
\end{array}
\right)=
\left(
\begin{array}{cccc}
 -q_{F^{Q_4}} & q_{F^{P^3}} & q_{F^{P^2}} & q_{F^{Q_1}} \\
 q_{F^{P^1}} & q_{F^{Q_2}} & q_{F^{Q_3}}& -q_{F^{P^4}} \\
\end{array}
\right)
\eeq
has the following dressing
\beq
\label{eq:3qubitdressedrelation}
\psi_{1ijk}=D^{(1)}_{ii'}D^{(2)}_{jj'}D^{(3)}_{kk'}\psi_{0i'j'k'},
\eeq
or equivalently,
\beq
|\psi_2\rangle =(D^{(1)} \otimes D^{(2)}  \otimes D^{(3)} )|\psi_1 \rangle ,
\eeq
with
\beq
D^{(i)}=\frac{1}{Y_i}
\left(
\begin{array}{cc}
X_i & -X_i^2-Y_i^2 \\
1 & -X_i
\end{array}
\right).
\eeq
This is the generalization of the relation \eqref{eq:relatedqubits} and shows that the pair of three qubit states are related by a moduli dependent $SL(2)^{3 \times}_U$ transformation. 
The formula \eqref{eq:Finv1} for $F$ is then valid for arbitrary scalar moduli provided that we use the dressed $R_i$ matrices of \eqref{eq:dressedRis} and the $K_{ab}$ matrices calculated from \eqref{eq:dressedfirstqubit} and \eqref{eq:3qubitdressedrelation} through \eqref{eq:fundcovariants}. 

Now we can also relate the charge matrix of an arbitrary solution through U-duality to an auxiliary charge matrix with trivial scalar asymptotics. Indeed, looking at \eqref{eq:cosetelement} it is manifest that we can remove the scalar hair from the coset element $\mathcal{V}$ by the U-duality transformation $\mathcal{V} \mapsto \mathcal{V}\mathcal{J}^{-1}$ where
\beq
\mathcal{J}=e^{-\frac{1}{2}\sum_i \log Y_i H_i}e^{-\sum_i X_i E_i},
\eeq
is the $SO(4,4)$ matrix corresponding to the $SL(2)^{\times 3}_U$ element
\bea
J_1\otimes J_2 \otimes J_3 \in SL(2)^{\times 3}_U, && J_i=\frac{1}{\sqrt{Y_i}}\left(
\begin{array}{cc}
1 & -X_i \\
0 & Y_i
\end{array}
\right).
\eea
Notice that due to \eqref{eq:Ucompensator} this transformation does not require a local compensator to go back to the Iwasawa gauge with $\mathcal{V}$. One may check with explicit computation that the asymptotic value of the coset element \eqref{eq:cosetMexp} indeed satisfies
\beq
\mathcal{J}^\# \mathcal{J} =\mathcal{M}^{(0)}.
\eeq
As a consequence, for any general charge matrix $Q$ we may define a duality transformed one
\beq
\label{eq:Qtilde}
\tilde Q= \mathcal{J} Q \mathcal{J}^{-1},
\eeq
which corresponds to a black hole with trivial moduli. Now we use the relations
\bea
D^{(i)}=
J_i^{-1}
\left(
\begin{array}{cc}
0 & -1 \\
1 & 0
\end{array}
\right) 
J_i, 
&&
R_i=J_i^{-1} \left(
\begin{array}{cc}
\Sigma_i & -\frac{\Xi_i}{Y_i} \\
-\frac{\Xi_i}{Y_i} & -\Sigma_i
\end{array}
\right)   J_i
\eea
to relate the auxiliary charges of $\tilde Q$ to the physical ones of $Q$. Following \ref{sec:actionofU} we deduce that the dyonic charge vectors $|\tilde \psi_{1,2}\rangle$ of $\tilde Q$ are expressed with the physical charges $|\psi_1 \rangle$ as
\bea
|\tilde \psi_1\rangle &=(J_1\otimes J_2 \otimes J_3)|\psi_1 \rangle, \\
|\tilde \psi_2\rangle &=(J_1\otimes J_2 \otimes J_3)(D^{(1)} \otimes D^{(2)}  \otimes D^{(3)} )|\psi_1 \rangle \\ &\equiv 
\left(
\begin{array}{cc}
0 & -1\\
1 & 0
\end{array}
\right)
\otimes
\left(
\begin{array}{cc}
0 & -1\\
1 & 0
\end{array}
\right)
\otimes
\left(
\begin{array}{cc}
0 & -1\\
1 & 0
\end{array}
\right)(J_1\otimes J_2 \otimes J_3)|\psi_1 \rangle.
\eea
We see that the vectors $|\tilde\psi_{1,2}\rangle$ are indeed related as in \eqref{eq:relatedqubits} which is valid only for the trivial moduli. For the scalar charges one has
\beq
\tilde R_i=J_i R_i J_i^{-1} \equiv \left(
\begin{array}{cc}
\Sigma_i & -\frac{\Xi_i}{Y_i} \\
-\frac{\Xi_i}{Y_i} & -\Sigma_i
\end{array}
\right).
\eeq
Since the $F$-invariant is blind to the transformation \eqref{eq:Qtilde} we obtained the result that for any asymptotics we may calculate $F$ by just using the formula for canonical  moduli with replacing the dyonic charges by $(J_1\otimes J_1 \otimes J_3)|\psi_1 \rangle$ and scaling the scalar charges as $\Xi_i \mapsto \frac{\Xi_i}{Y_i}$, i.e.
\beq
\label{eq:Fdressing}
F(X_i,Y_i,|\psi_1\rangle,\Sigma_i,\Xi_i)=F(X_i=0,Y_i=1,(J_1\otimes J_2 \otimes J_3)|\psi_1\rangle,\Sigma_i,\frac{\Xi_i}{Y_i}).
\eeq

\section{Relations between scalar charges and physical charges}
\label{sec:scalcharg}

Our formula is entirely in terms of the asymptotic charges of the black hole and is manifestly invariant under U-duality and permutation of scalars. However, it does contain explicitly the scalar charges $\Sigma_i$ and $\Xi_i$ which are not independent of $M$, $N$, $Q_I$ and $P^I$. A formula entirely in terms of the physical charges would require solving for the functions $\Sigma_i(M,N,Q,P)$, $\Xi_i(M,N,Q,P)$. In this section, we provide constraints that these functions must satisfy and solve them for some special cases. We illustrate on the example of the four electric charge Cveti\v c-Youm black hole that in general it is not possible to give the $F$-invariant in terms of radicals of the physical charges.

We start by describing the constraint equations. Recall, that the charge matrix $Q$ of \eqref{eq:chargematrix} tranforms as a four qubit state (see \eqref{eq:fourqubitSLOCC}) under the action of the $SL(2)^{4\times}$ spanned by the second line of \eqref{eq:splitsummary}. The charge-up matrix $h$ of e.q. \eqref{eq:chargeup} is an element of this $SL(2)^{4\times}$. It is known\cite{verstraete,levay4qubit} that this 16 dimensional representation admits four algebraically independent continous invariants. All of these can be checked to be proportional to some power of the combination $(m^2+n^2)$ of the seed parameters. It follows that they provide 3 independent polynomial equations among the 16 asymptotic charges in $Q$. Let us describe a convenient and simple way of obtaining these equations. Consider the characteristic polynomial of the charge matrix $Q$:
\beq
\label{eq:charpol}
p(\lambda)=\det(\lambda I-Q).
\eeq
It is clear that the characteristic polynomial is invariant under the charge up operation and hence we have
\beq
p(\lambda)=p_0(\lambda),
\eeq
where the characteristic polynomial for the seed solution is
\bea
p_0(\lambda) &=\det(\lambda I-Q_{KTN})\\
&=\lambda ^4 \left(\lambda ^2-4 m^2-4 n^2\right)^2\\
&=\lambda ^4 \left(\lambda ^2-\frac{1}{4}\text{Tr}Q^2\right)^2,
\eea
where we have used the invariance of $\text{Tr}Q^2$ to express $m^2+n^2$ in terms of asymptotic charges. The two polynomials agree iff all of their coefficients agree hence we have the following 9 polynomial equations
\beq
\label{eq:constraints}
\frac{\dd^k}{\dd \lambda^k}(p(\lambda)-p_0(\lambda))|_{\lambda=0}=0, \;\; k=0,...,8.
\eeq
One can check that for $k=1,3,5,6,7,8$ these are trivially satisfied and hence we are left with 3 equations. These 3 equations have linearly independent gradients in the scalar charges which indicates a three dimensional solution set. However, this does not tell us anything about the set of real solutions. One can easily check that as one approaches the seed solution by taking $Q_I\rightarrow 0$ and $P^I\rightarrow 0$, there is only one real root satisfying the consistency requirement $\Sigma_i \rightarrow 0$. This shows that it is possible that the three equations are enough to determine the scalar charges uniquely. To say something more precise about this one would need to determine at least the real dimension of the semialgebraic set defined by these equations, but to our knowledge, there is no method to do this to date. Instead, we provide solutions to \eqref{eq:constraints} for some special charge vectors, where the $F$-invariant is known explicilty, and hence we can compare our results with the existing literature. 

Before doing so, we comment on what happens with these constraints when one considers black holes with non-trivial asymptotic moduli. In this case one can just replace $Q$ in \eqref{eq:charpol} with the auxiliary charge vector $\tilde Q$, as readily seen from \eqref{eq:Qtilde}. This shows that whatever expressions $\Sigma_i=f_i(M,|\psi_1\rangle)$, $\Xi_i=g_i(M,|\psi_1\rangle)$ we find for trivial moduli by solving \eqref{eq:constraints}, we can safely use them for non-trivial moduli as $\Sigma_i=f_i(M,(J_1\otimes J_2 \otimes J_3)|\psi_1\rangle)$ and $\Xi_i=Y_i g_i(M,(J_1\otimes J_2 \otimes J_3)|\psi_1\rangle)$. Combine this with \eqref{eq:Fdressing} to get the expected result
\beq
\label{eq:Fdressing2}
F(X_i,Y_i,|\psi_1\rangle)=F(X_i=0,Y_i=1,(J_1\otimes J_2 \otimes J_3)|\psi_1\rangle),
\eeq
which is then guaranteed to be valid as long as we can use \eqref{eq:constraints} to solve for the scalar charges. This is the case for the first two of the following examples.

\subsection{Klauza-Klein black hole}

The Klauza-Klein black hole\cite{rasheed1995rotating} is obtained by setting $Q_2=Q_3=Q_4=0$ and $P^2=P^3=P^4=0$ and $N=0$. From the parametrization \eqref{eq:deltaderiv1}-\eqref{eq:deltaderiv2}  we can deduce that $\Xi_i=0$ and $\Sigma_2=\Sigma_3=-\Sigma_1$ in this case. We can put this into \eqref{eq:constraints} as an ansatz and observe that it automaticaly solves two equations. The third equation reads as
\beq
\Sigma_1 \left(8 M^2+(P^1)^2+Q_1^2-2 \Sigma_1^2\right)=2 M \left((P^1)^2-Q_1^2\right).
\eeq
As expected, there is only one root that vanishes for zero charges. For this root we have found numerical agreement between our formula
\beq
F=\left( M^2-\frac{1}{4}(P^1)^2\right) \left( M^2-\frac{1}{4}Q_1^2\right)+\frac{1}{8} M \Sigma_1 ((P^1)^2-Q_1^2) -\frac{1}{16}\Sigma_1^4,
\eeq
 for the F-invarant and the complicated expression presented in \cite{compere2} for the Klauza-Klein black hole. We may further specialize by setting $P^1=0$. In this case the above equation factorizes as
\beq
(2 M+\Sigma_1) \left(Q_1^2+4 M \Sigma_1-2 \Sigma_1^2\right)=0.
\eeq
The physical root is $\Sigma_1=M-\sqrt{M^2+\frac{1}{2}Q_1^2}$. Upon substituing this into the formula for the $F$-invariant we get
\beq
F=\frac{1}{64} \bigg(  32 M^4-40 M^2 Q_1^2-Q_1^4+4M (4 M^2+ 2Q_1^2)^{\frac{3}{2}}\bigg),
\eeq
in complete agreement with \cite{compere2}.

\subsection{$-i X^0 X^1$ supergravity black hole}

Now let us consider the axion-dilaton black hole of \cite{lozano2000general}. We set the electric and magnetic charges pairwise equal $Q_1=Q_4$, $Q_2=Q_3$, $P^1=P^4$ and $P^2=P^3$. We also set the NUT charge to zero. From \eqref{eq:deltaderiv1}-\eqref{eq:deltaderiv2} we observe that in this case we have $\Sigma_2=\Sigma_3=\Xi_2=\Xi_3=0$. Using this as an ansatz in \eqref{eq:constraints} we are left with a single equation
\bea
2 (P^1)^2 & \left(2 M \Sigma_1+(P^2)^2-Q_1^2-Q_2^2\right)+2 Q_2 \left(4 M \Xi_1 (P^2)+2 M Q_2 \Sigma_1+Q_1^2 Q_2\right)\\
&=4 M^2 \Xi_1^2+4 M^2 \Sigma_1^2+(P^1) (8 M \Xi_1 Q_1-8 (P^2) Q_1 Q_2)\\
&+2 (P^2)^2 \left(2 M \Sigma_1+Q_1^2+Q_2^2\right)+4 M Q_1^2 \Sigma_1+(P^1)^4+(P^2)^4+Q_1^4+Q_2^4,
\eea
which admits a single real solution
\bea
\Sigma_1 &=\frac{ (P^1)^2-(P^2)^2-Q_1^2+Q_2^2}{2 M},\\
\Xi_1 &=\frac{P^2 Q_2-P^1 Q_1}{M}.
\eea
This agrees with the axion-dilaton charge obtained in \cite{lozano2000general}
\beq
\Upsilon=i(\Xi_1-i \Sigma_1)=-\frac{(Q_1+i P^1)^2+(-P^2+i Q_2)^2}{2 M}.
\eeq
Upon inserting this into our formula \eqref{eq:Finv1} for the $F$-invariant we obtain
\beq
F=\frac{1}{16} \left(4 M^2-(P^1-P^2)^2-(Q_1-Q_2)^2\right) \left(4 M^2-(P^1+P^2)^2-(Q_1+Q_2)^2\right),
\eeq
in complete agreement with \cite{compere2,lozano2000general}. We note here that when the single modulus of this model is turned on we have to replace the charges according to \eqref{eq:Fdressing2}. Explicitly, this leads to the following replacement rule
\bea
\left( \begin{array}{c}
P^1 \\
Q_1
\end{array}
\right) \mapsto
\frac{1}{\sqrt{Y_1}}\left(
\begin{array}{cc}
1 & -X_1 \\
0 & Y_1
\end{array}
\right)
\left( \begin{array}{c}
P^1 \\
Q_1
\end{array}
\right), 
&&
\left( \begin{array}{c}
Q_2 \\
-P^2
\end{array}
\right) \mapsto
\frac{1}{\sqrt{Y_1}}\left(
\begin{array}{cc}
1 & -X_1 \\
0 & Y_1
\end{array}
\right)
\left( \begin{array}{c}
Q_2 \\
-P^2
\end{array}
\right),
\eea
which, again, agrees with \cite{lozano2000general}.

\subsection{Dilute gas limit}

We can recover the dilute gas limit of \cite{larsen} as well. In this limit, we have the following constraint among the magnetic charges
\beq
P^1+P^2+P^3+P^4=0.
\eeq
Provided that this is true, all three equations of \eqref{eq:constraints} can be solved \textit{exactly} by setting
\bea
\label{eq:dilutesc}
\Sigma_1 &=-2 M+Q_2+Q_3, && \Sigma_2 =& 2M-Q_2-Q_4, && \Sigma_3 &= 2M-Q_3-Q_4,\\
\Xi_1 &=P^2+P^3, && \Xi_2 =&-P^2-P^4, && \Xi_3 &=-P^3-P^4.
\eea
These roots cannot be physical for all values of the charges as for vanishing charges we must have $\Sigma_i=0$ . However, in the dilute gas limit, the charges are large and hence this requirement is outside of the region of validity. Define the excitation energy as $\delta M=M-M_{BPS}=M-\frac{1}{4}(Q_1+Q_2+Q_3+Q_4)$. We obtain the dilute gas limit of F by substituing \eqref{eq:dilutesc} into \eqref{eq:Finv1} and scaling $Q_i\rightarrow \mu^2 Q_i$, $i=2,3,4$ and $P^I \rightarrow \mu P^I$, while keeping $\delta M$ fixed. Upon $\mu \rightarrow \infty$ the leading order $\mu^8$ in F vanishes. The next to leading order contribution is the coefficient of $\mu^6$ which is
\beq
F_0=\frac{1}{2}\delta M Q_2 Q_3 Q_4,
\eeq
which agrees with the result of \cite{larsen}. 

\subsection{Four charge Cveti\v c-Youm black hole}

Here we set all the magnetic charges and the NUT charge to zero but allow for arbitrary electric charges\cite{cvetivc1996entropy}. We have all $\Xi_i=0$ but we still need to solve for all the $\Sigma_i$. Instead of trying to solve the constraint equations we can simply use that as all $\gamma_I=0$, the equations for the $\delta_I$ derivatives \eqref{eq:deltaderiv1}-\eqref{eq:deltaderiv2} give the full Jacobian for the change of variables from boost parameters to asymptotic charges. The best thing to do is to write differential equations for the functions $Q_I(2M,\Sigma_i)$ because these are remarkably easily solved. The solution is such that
\beq
\sqrt{4 Q_I^2+C_I}=\frac{\partial Q_I}{\partial \delta_I},
\eeq
where the right hand side is understood to be given through \eqref{eq:deltaderivQ}. The $C_I$ are constants of integration. Comparing with the actual parametrization reveals that $C_I=4m$, $I=1,...,4$. The solution for the scalar charges is then easily obtained to be
\bea
\Sigma_1 &= -\frac{1}{2} \sqrt{m^2+Q_1^2}+\frac{1}{2} \sqrt{m^2+Q_2^2}+\frac{1}{2} \sqrt{m^2+Q_3^2}-\frac{1}{2} \sqrt{m^2+Q_4^2},\\
\Sigma_2 &= \frac{1}{2} \sqrt{m^2+Q_1^2}-\frac{1}{2} \sqrt{m^2+Q_2^2}+\frac{1}{2} \sqrt{m^2+Q_3^2}-\frac{1}{2} \sqrt{m^2+Q_4^2},\\
\Sigma_3 &= \frac{1}{2} \sqrt{m^2+Q_1^2}+\frac{1}{2} \sqrt{m^2+Q_2^2}-\frac{1}{2} \sqrt{m^2+Q_3^2}-\frac{1}{2} \sqrt{m^2+Q_4^2}.
\eea
Plugging these expressions into the formula \eqref{eq:Finv1} for the $F$-invariant we recover the expression given in \cite{compere2}. This expression still depends on the seed parameter $m$. It is determined in terms of the physical charges through the equation
\beq
\label{eq:masscveticyoum}
M=\frac{1}{4}\left(  \sqrt{m^2+Q_1^2}+ \sqrt{m^2+Q_2^2}+\sqrt{m^2+Q_3^2}+ \sqrt{m^2+Q_4^2}\right).
\eeq
We happily acknowledge that the right hand side is greater than $\frac{1}{4}\left( Q_1+Q_2+Q_3+Q_4 \right)$, and hence the requirement of solvability is $M\geq M_{BPS}$. Note that this example illustrates that it is in general not possible to express the $F$-invariant in terms of radicals of the physical charges. Indeed, one may rewrite \eqref{eq:masscveticyoum} as a system of five polynomial equations as $M=\frac{1}{4}\sum_{I=1}^4 x_I$ and $x_I^2=m^2+Q_I^2$ for the five variables $m^2$ and $x_I$. Then one may use some algorithm to cast this system into regular chains. The first element of the chain can be chosen to depend only on $m^2$ and then to aquire $m$ we need to consider only this equation and forget about the others\footnote{Note that not all of the roots of this equation are solutions to \eqref{eq:masscveticyoum}.}. We do not present this equation here due to its length but it is a general, fifth order polynomial equation for $m^2$. Then, due to the Abel-Ruffini theorem, one cannot have an expression for $m$ in terms of radicals of the coefficients.

\section{Generalization for $\mathcal{N}=8$ supergravity}
\label{sec:generalize}

We have seen that in the STU case the charge matrix is an element of $\mathfrak{so}(4,4)$ and the U-duality group $SL(2)^{\times 3}_U\subset SO(4,4)$ acts on it through the adjoint representation of $SO(4,4)$. This representation decomposes as $28=1\oplus 1\oplus 1\oplus 9\oplus 8\oplus 8$. There is a general way of constructing invariants on $9\oplus 8\oplus 8$ which allowed us to identify the $F$-invariant. In the $\mathcal{N}=8$ case the 3d coset model is $E_{8(8)}/SO^*(16)$ and the U-duality group is $E_{7(7)}$. We expect that in this case the asymptotic charges of the black hole parametrize a Lie-algebra element $Q\in \mathfrak{e}_8$ and the U-duality group  $E_{7(7)}\subset E_{8(8)}$ just acts by the adjoint action of $E_{8(8)}$. This representation decomposes as $248=1\oplus 1\oplus 1\oplus 133\oplus 56\oplus 56$ and hence the relevant representation space is $133\oplus 56 \oplus 56$ with $56$ replacing three qubit states containing dyonic charges and $133$ replacing $9$ containing the 70 scalar charges \footnote{Recall that in the STU case $9$ contained 6 scalar charges, this is just an artifact of fixing the scalar asymptotics.}. The moment map from pairs of $56$ to $133$ can be formulated. 
The construction goes as follows. There is an $E_{7(7)}$ invariant antisymmetric bilinear form on 56, let us denote this by $\langle .,.\rangle$. We may define an $\mathfrak{e}_7$ element $T_{\Psi_1 \Psi_2}$ associated to the pair $\Psi_1,\Psi_2\in 56$ by demanding
\beq
\label{eq:momentdef}
\kappa(T,T_{\Psi_1 \Psi_2})=\langle \Psi_1,T \Psi_2 \rangle, \;\; \forall T\in \mathfrak{e}_7.
\eeq
Here, $\kappa$ is the Killing form on $\mathfrak{e}_7$. Using $56\cong \wedge^2 \mathbb{C}^8\oplus \wedge^2 (\mathbb{C}^8)^*$ we may parametrize $\Psi_a$, $a=1,2$ with a pair of antisymmetric $8\times 8$ matrixes:
\beq
\Psi_a=\big((x^{(a)})^{ij},y^{(a)}_{ij}\big), \;\;(x^{(a)})^{ij}=-(x^{(a)})^{ji}, \;\; y^{(a)}_{ij}=-y^{(a)}_{ji}
\eeq
and using $\mathfrak{e}_7\cong \mathfrak{sl}_8 \oplus \wedge^4 \mathbb{C}^8$ we can parametrize the generators as
\beq
T=({\Lambda^i}_j,\Sigma_{ijkl}), \;\; {\Lambda^i}_i=0,
\eeq
and $\Sigma_{ijkl}$ totaly antisymmetric. We refer to the appendix of \cite{julia1} for the commutation relations, the action of $\mathfrak{e}_7$ on $56$ and the Killing form in terms of this parametrization. The invariant bilinear product reads as
\beq
\langle \Psi_1,\Psi_2 \rangle=(x^{(1)})^{ij}y^{(2)}_{ij}-(x^{(2)})^{ij}y^{(1)}_{ij}.
\eeq
Then, using the definition \eqref{eq:momentdef}, a short excercise reveals the explicit form of the moment map to be
\bea
T_{\Psi_a \Psi_b}&=\big({(\Lambda_{(ab)})^i}_j,(\Sigma_{(ab)})_{ijkl}\big),\\
{(\Lambda_{(ab)})^i}_j &=-\frac{1}{6}\big(  (x^{(a)})^{in}y^{(b)}_{jn} + (x^{(b)})^{in}y^{(a)}_{jn} \big)+\frac{1}{48}\big( (x^{(a)})^{nm}y^{(b)}_{nm} + (x^{(b)})^{nm}y^{(a)}_{nm}\big){\delta^i}_j,\\
(\Sigma_{(ab)})_{ijkl} &=\frac{1}{48}\big( \epsilon_{ijklmnop}(x^{(a)})^{mn}(x^{(b)})^{op}-  y^{(a)}_{[ij}y^{(b)}_{kl]} \big).
\eea
Note that in this formalism we have the Cartan-Cremmer-Julia invariant expressed as
\bea
I_4 &\equiv \frac{1}{2}\langle \Psi,T_{\Psi \Psi} \Psi \rangle \\
&= x^{ij}x^{kl}y_{ik}y_{jl}-\frac{1}{4}(x^{ij}y_{ij})^2+\frac{1}{96}\big( \epsilon_{ijklmnop} x^{ij} x^{kl} x^{mn} x^{op}+\epsilon^{ijklmnop}y_{ij}y_{kl}y_{mn}y_{op} \big).
\eea
The conventions of \cite{compere2} are such that $I_4=4 \diamondsuit$ and $\diamondsuit$ is the one that reduces to $\Delta$ of \eqref{eq:cayley} for the STU duality frame. The dyonic charges parametrize $\Psi_1,\Psi_2\in 56$ generalizing $\psi_1,\psi_2$. The STU charges \eqref{eq:pairofthreequbits} sit inside this $\Psi_1$ and $\Psi_2$ as
\bea
\label{eq:8of56}
\left(
\begin{array}{cccc}
y^{(2)}_{12} & (x^{(2)})^{34} & (x^{(2)})^{56}  & y^{(2)}_{78}\\
(x^{(2)})^{78} & y^{(2)}_{56} & y^{(2)}_{34} &  (x^{(2)})^{12}\\ \hline
y^{(1)}_{12} & (x^{(1)})^{34} & (x^{(1)})^{56}  & y^{(1)}_{78}\\
(x^{(1)})^{78} & y^{(1)}_{56} & y^{(1)}_{34} &  (x^{(1)})^{12}
\end{array}
\right) &=\frac{1}{\sqrt{2}}
\left(
\begin{array}{cccc}
\psi_{0000} & \psi_{0001} & \psi_{0010} & \psi_{0011}\\
\psi_{0100} & \psi_{0101} & \psi_{0110} & \psi_{0111}\\ \hline
\psi_{1000} & \psi_{1001} & \psi_{1010} & \psi_{1011}\\
\psi_{1100} & \psi_{1101} & \psi_{1110} & \psi_{1111}
\end{array}
\right)\\
&=\frac{1}{\sqrt{2}}
\left(
\begin{array}{cccc}
-P^4 & -Q_3 & -Q_2 & P^1\\
-Q_1 & P^2 & P^3 & Q_4\\ \hline
-Q_4 & P^3 & P^2 & Q_1\\
P^1 & Q_2 & Q_3 & -P^4
\end{array}
\right),
\eea
with all the remaining $(x^{(a)})^{ij}$ and $y^{(a)}_{ij}$ vanishing. See \cite{kalloshlinde,duff5} for details. One imediately verifies that the relation to the fermionic inner product of e.q. \eqref{eq:fermioninnerprod} is simply
\beq
\langle \Psi_1, \Psi_2 \rangle = (P_{\psi_1},P_{\psi_2}),
\eeq
and that $\diamondsuit$ reduces to $\Delta$. As the next step, we may parametrize an element $R\in \mathfrak{e}_7$ with 70 scalar charges. Denote the corresponding $56 \times 56$ matrix in the adjoint representation with $\mathcal{R}$. In the STU duality frame $\mathcal{R}$ should reduce to $\hat R$ of \eqref{eq:Rfundamental} on the eight dimensional subspace of 56, where the pair of \eqref{eq:8of56} lives, and zero everywhere else. Then, the $F$ invariant \eqref{eq:Finv2} can be written in a manifestly $E_7$ invariant way as:
\bea
\label{eq:e7Finv}
F&=M^4+M^2 N^2+\frac{ M^2}{4}\langle \Psi_1,\Psi_2\rangle+\frac{M }{24}\langle \Psi_1,\mathcal{R} \Psi_2\rangle+\frac{N}{24}\langle \Psi_1,\mathcal{R} \Psi_1 \rangle\\
&+\frac{N^2}{96}\text{Tr}(\mathcal{R}^2)-\frac{1}{192}\langle \Psi_1,(T_{\Psi_1 \Psi_2}-T_{\Psi_2 \Psi_2} ) \Psi_1 \rangle\\
&+\frac{1}{1536}\left(\frac{1}{8}(\text{Tr}\mathcal{ R}^2)^2-\text{Tr}(\mathcal{ R}^4)\right).
\eea

\textit{Note added on \today}: After the completion of this work we have learned that for any generator $T$ of $E_7$ represented in $56$ the invariant $\text{Tr}T^4$ satisfies
\beq
\text{Tr}T^4=\frac{1}{24}(\text{Tr}T^2)^2,
\eeq
and hence it is not an independent invariant. It follows that the last piece of \eqref{eq:e7Finv} is not correct and greater care is needed for the embedding of the STU scalar charges. This was subsequently done in \cite{lekeu} where the correct formula for the $E_7$ invariant version of $F$ can be found. 

\section{Conclusions}


In this work we managed to express the $F$-invariant of Chow and Comp\` ere, and hence the Bekenstein-Hawking entropy of general asymptotically flat (or Taub-NUT), non-extremal black holes admitted by the STU model in terms of 16 asymptotic charges. These are not all independent: six scalar charges are functions of the mass, NUT charge, and eight dyonic charges. However, the expression for $F$ with the scalar charges being explicit makes the U-duality invariance manifest and allowed us to conjecture the generalization \eqref{eq:e7Finv} of the $F$-invariant to the $E_{7(7)}$ invariant case. We have argued that a formula in terms of only the physical charges requires one to find the real solutions of a system of polynomial equations. We have solved these equations for some known special cases and recovered the expected expressions.

An important question left open is whether equations \eqref{eq:constraints}, together with the condition of reality, are enough to determine the scalar charges uniquely or one needs some additional constraints. Also, it would be very interesting to see a microscopic origin of this entropy formula. There are several results along these lines including near-extremal black holes\cite{witten,horowitz1,horowitz2,horowitz3}, neutral, non-extremal ones\cite{alejandra} and the recently constructed dilute gas limit of the general non-extreme STU black holes\cite{larsen}. Another interesting problem would be to find the uplift of the formula for 5 dimensional finite temperature black holes and black rings written as a qubic invariant of $E_{6(6)}$\cite{strominger}. A way of finding this formula would probably be to recast our expression for the $F$-invariant in the language of Freudenthal triple systems\cite{krutelevich} and write it in terms of elements of the corresponding cubic Jordan algebra\cite{duff3}.

Finally, as adding a possible new twist to the black hole/qubit correspondence, we note that the difference of the inner and outer horizon radius is\cite{compere2}
\beq
r_+ -r_- = 2\sqrt{m^2+n^2-a^2} =\frac{1}{2}\sqrt{\text{Tr}Q^2\left( 1-\frac{J^2}{F}\right)},
\eeq
which measures extremality, and is U-duality invariant as expected. It is known that the quantity $\text{Tr}Q^2$ can be reinterpreted though \eqref{eq:Qexpand} and \eqref{eq:sigmaqubits} as the quadratic four qubit entanglement measure\cite{levay4qubit}. Then, so called nilpotent states with $\text{Tr}Q^2=0$ allways correspond to extremal black holes and they can be classified in the language of four qubit entanglement\cite{duff1}. However, we see that there are extremal black holes corresponding to semisimple charge vectors: these are the extremal, fast rotating black holes\cite{compere2} with $F=J^2$. It would be interesting to see if these black holes relate to the entanglement properties of semisimple four qubit states.

\section*{Acknowledgments}

I am grateful for Alejandra Castro for bringing my attention to the problem. 
I would like to thank Alejandra Castro, Gary Horowitz, P\' eter L\' evay and Joe Polchinski for reading the manuscript and providing valuable feedback. I am grateful for Geoffrey Comp\' ere, Victor Lekeu and Mario Trigiante for useful correspondence. I would like to thank the Gordon and Betty Moore Foundation for financial support. This research was supported in part by the National Science Foundation under Grant No. NSF PHY11-25915.

\appendix

\section{Different ways of splitting $\mathfrak{so}(4,4)$ as $\mathfrak{sl}^{\times 4}_2\oplus (2,2,2,2)$}
\label{sec:liealg}

We define the group $SO(4,4)$ as the set of $8\times 8$ matrices $O$ keeping the bilinear form
\beq
G=\left(
\begin{array}{cc}
 0 & I\\
 I & 0
\end{array}
\right),
\eeq
i.e. $OGO^T=G$. Here, $I$ is the $4\times 4$ identity matrix. The Lie algebra $\mathfrak{so}(4,4)$ is spanned by matrices $T$ satisfying $TG+GT^T=0$.
We use the same parametrization of these matrices as in \cite{compere2}. We denote by $E_{i j}$ the $8 \times 8$ matrix with 1 in the $(i, j)$ component and zeros everywhere else. The Cartan generators are given by
\bea
H_0 &=E_{33}+E_{44}-E_{77}-E_{88}, && H_1 &=E_{33}-E_{44}-E_{77}+E_{88},\\
H_2 & = E_{1 1} + E_{2 2} - E_{5 5} - E_{6 6} , && H_3 & = E_{1 1} - E_{2 2} - E_{5 5} + E_{6 6} ,
\eea
while the roots are parametrized as
\bea
E_0 & = E_{4 7} - E_{3 8} ,& E_1 & = E_{8 7} - E_{3 4} ,& E_2 & = E_{2 5} - E_{1 6} ,& E_3 & = E_{6 5} - E_{1 2} , \\
E^{Q_1} & = E_{4 5} - E_{1 8} , & E^{Q_2} & = E_{3 2} - E_{6 7} , & E^{Q_3} & = E_{3 6} - E_{2 7} , & E^{Q_4} & = E_{4 1} - E_{5 8} , \\
E^{P^1} & = E_{5 7} - E_{3 1} , & E^{P^2} & = E_{4 6} - E_{2 8} , & E^{P^3} & = E_{4 2} - E_{6 8} , & E^{P^4} & = E_{1 7} - E_{3 5} , \\
F_0 & = E_{7 4} - E_{8 3} ,& F_1 & = E_{7 8} - E_{4 3} ,& F_2 & = E_{5 2} - E_{6 1} ,& F_3 & = E_{5 6} - E_{2 1} , \\
F^{Q_1} & = E_{5 4} - E_{8 1} , & F^{Q_2} & = E_{2 3} - E_{7 6} , & F^{Q_3} & = E_{6 3} - E_{7 2} , & F^{Q_4} & = E_{1 4} - E_{8 5} , \\
F^{P^1} & = E_{7 5} - E_{1 3} , & F^{P^2} & = E_{6 4} - E_{8 2} ,& F^{P^3} & = E_{2 4} - E_{8 6} , & F^{P^4} & = E_{7 1} - E_{5 3} .
\eea

\subsection{U-duality split}
\label{sec:Udualitysplit}

It is easy to see that the generators $H_\Lambda$, $E_\Lambda$, $F_\Lambda$, $\Lambda=0,1,2,3$ form four commuting $\mathfrak{sl}_2$ algebras:
\bea
\label{eq:firstSL2}
\; [H_\Lambda,E_\Lambda]=2E_\Lambda, && [H_\Lambda,F_\Lambda]=-2F_\Lambda, && [E_\Lambda,F_\Lambda]=H_\Lambda.
\eea
The remaining $16$ generators $E^{P^I}$, $E^{Q_I}$, $F^{P^I}$, $F^{Q_I}$ form the fundamental $(2,2,2,2)_U$ representation of this $(\mathfrak{sl}^{\times 4}_2)_U$ algebra under the adjoint action. 

A vector of this representation can nicely be described by the 16 amplitudes $\psi_{ijkl}$ of a four qubit state 
\beq
|\psi\rangle =\sum_{i,j,k,l=0}^1 \psi_{ijkl}|ijkl\rangle,
\eeq
transforming as
\beq
\label{eq:fourqubitSLOCC}
\psi_{ijkl}\mapsto {(S_0)_i}^{i'} {(S_1)_j}^{j'} {(S_2)_k}^{k'} {(S_3)_l}^{l'}\psi_{i'j'k'l'}, \;\; S_0\otimes S_1\otimes S_2\otimes S_3 \in SL(2)^{\times 4},
\eeq
under the action of the group $SL(2)^{\times 4}$ generated by the algebra \eqref{eq:firstSL2}. In terms of the Lie algebra generators one writes this vector as
\bea
\label{eq:Udual4qubit}
\Psi &= \psi_{0000} E^{P^4}+\psi_{0001}E^{Q_3}+\psi_{0010}E^{Q_2}-\psi_{0011}E^{P^1}\\
     &+\psi_{0100}E^{Q_1}-\psi_{0101}E^{P^2}-\psi_{0110}E^{P^3}-\psi_{0111}E^{Q_4}\\
     &-\psi_{1000}F^{Q_1}+\psi_{1001}F^{P^3}+\psi_{1010}F^{P^2}+\psi_{1011}F^{Q_1}\\
     &+\psi_{1100}F^{P^1}+\psi_{1101}F^{Q_2}+\psi_{1110}F^{Q_3}-\psi_{1111}F^{P_4},
\eea
transforming as \eqref{eq:fourqubitSLOCC} under $\Psi \mapsto g \Psi g^{-1}$.

\subsection{Sigma model split}
\label{sec:secondsplit}

We can realize the split $\mathfrak{sl}^{\times 4}_2\oplus (2,2,2,2)$ in a different way suited to writing the timelike reduced STU action \eqref{eq:3daction} as a sigma model on $SO(4,4)/SL(2)^{\times 4}$. We may introduce the symmetric bilinear form
\beq
\eta=\text{diag}(-1,-1,1,1,-1,-1,1,1),
\eeq
and look for the subgroup $SO(2,2)\times SO(2,2)\cong SL(2)^{\times 4}$ keeping this fixed. This subgroup is generated by the $-1$ eigenspace of the involution
\beq
\label{eq:sharpdef}
T^\#=\eta T^T \eta.
\eeq
This eigenspace is spanned by the 12 generators
\bea
k_\Lambda & = E_\Lambda - F_\Lambda , && k^{Q_I} & = E^{Q_I} + F^{Q_I} , && k^{P^I} & = E^{P^I} + F^{P^I} ,
\eea
while the $+1$ eigenspace is spanned by
\bea
\label{eq:sigmafourqubit}
H_\Lambda, && p_\Lambda & = E_\Lambda + F_\Lambda , && p^{Q_I} & = E^{Q_I} - F^{Q_I} , && p^{P^I} & = E^{P^I} - F^{P^I} .
\eea
To see that this indeed realizes a $\mathfrak{sl}^{\times 4}_2\oplus (2,2,2,2)$ split define\cite{virmani,vernocke}
\bea
\tilde H_1&=1/2(-k^{Q_4}-k^{Q_1}-k^{Q_2}-k^{Q_3}),\\
\tilde H_2&=1/2(k^{Q_4}+k^{Q_1}-k^{Q_2}-k^{Q_3}),\\
\tilde H_3&=1/2(k^{Q_4}-k^{Q_1}+k^{Q_2}-k^{Q_3}),\\
\tilde H_4&=1/2(k^{Q_4}-k^{Q_1}-k^{Q_2}+k^{Q_3}),\\
\tilde E_1&=1/4(-k_0+k_1+k_2+k_3+k^{P^1}+k^{P^2}+k^{P^3}+k^{P^4}),\\
\tilde E_2&=1/4(k_0-k_1+k_2+k_3+k^{P^1}-k^{P^2}-k^{P^3}+k^{P^4}),\\
\tilde E_3&=1/4(k_0+k_1-k_2+k_3-k^{P^1}+k^{P^2}-k^{P^3}+k^{P^4}),\\
\tilde E_4&=1/4(k_0+k_1+k_2-k_3-k^{P^1}-k^{P^2}+k^{P^3}+k^{P^4}),\\
\tilde F_1&=1/4(k_0-k_1-k_2-k_3+k^{P^1}+k^{P^2}+k^{P^3}+k^{P^4}),\\
\tilde F_2&=1/4(-k_0+k_1-k_2-k_3+k^{P^1}-k^{P^2}-k^{P^3}+k^{P^4}),\\
\tilde F_3&=1/4(-k_0-k_1+k_2-k_3-k^{P^1}+k^{P^2}-k^{P^3}+k^{P^4}),\\
\tilde F_4&=1/4(-k_0-k_1-k_2+k_3-k^{P^1}-k^{P^2}+k^{P^3}+k^{P^4}).
\eea
One easily verifies that
\bea
\label{eq:secondSL2}
\; [\tilde H_J,\tilde E_J]=2\tilde E_J, && [\tilde H_J,\tilde F_J]=-2\tilde F_J, && [\tilde E_J,\tilde F_J]=\tilde H_J, 
\eea
with $J=1,..,4$ and all other commutators vanishing. We can write an element of the $+1$ eigenspace of $^\#$ in terms of four qubit amplitudes $\psi_{ijkl}$ transforming as in \eqref{eq:fourqubitSLOCC} under this new $SL(2)^{\times 4}$. It reads explicitly as
\beq
\tilde \Psi=\tilde \Psi_{H_\Lambda}H_\Lambda + \tilde \Psi_{p_\Lambda}p_\Lambda +\tilde \Psi_{Q_I}p^{Q_I}+\tilde \Psi_{P^I}p^{P^I},
\eeq
where
\bea
\label{eq:sigmaqubits}
\tilde \Psi_{H_0} &=\psi _{0001}+\psi _{0010}+\psi _{0100}-\psi _{0111}-\psi _{1000}+\psi _{1011}+\psi _{1101}+\psi _{1110},\\
\tilde \Psi_{H_1}&=\psi _{0001}+\psi _{0010}-\psi _{0100}+\psi _{0111}+\psi _{1000}-\psi _{1011}+\psi _{1101}+\psi _{1110},\\
\tilde \Psi_{H_2}&=\psi _{0001}-\psi _{0010}+\psi _{0100}+\psi _{0111}+\psi _{1000}+\psi _{1011}-\psi _{1101}+\psi _{1110},\\
\tilde \Psi_{H_3}&=-\psi _{0001}+\psi _{0010}+\psi _{0100}+\psi _{0111}+\psi _{1000}+\psi _{1011}+\psi _{1101}-\psi _{1110},\\
\tilde \Psi_{p_0}&=-\psi _{0000}+\psi _{0011}+\psi _{0101}+\psi _{0110}-\psi _{1001}-\psi _{1010}-\psi _{1100}+\psi _{1111},\\
\tilde \Psi_{p_1}&=-\psi _{0000}+\psi _{0011}-\psi _{0101}-\psi _{0110}+\psi _{1001}+\psi _{1010}-\psi _{1100}+\psi _{1111},\\
\tilde \Psi_{p_2}&=-\psi _{0000}-\psi _{0011}+\psi _{0101}-\psi _{0110}+\psi _{1001}-\psi _{1010}+\psi _{1100}+\psi _{1111},\\
\tilde \Psi_{p_3}&=-\psi _{0000}-\psi _{0011}-\psi _{0101}+\psi _{0110}-\psi _{1001}+\psi _{1010}+\psi _{1100}+\psi _{1111},\\
\tilde \Psi_{Q_1}&=2 \left(\psi _{0100}-\psi _{1011}\right),\\
\tilde \Psi_{Q_2}&=2 \left(\psi _{0010}-\psi _{1101}\right),\\
\tilde \Psi_{Q_3}&=2 \left(\psi _{0001}-\psi _{1110}\right),\\
\tilde \Psi_{Q_4}&=2 (\psi _{1000}- \psi _{0111}),\\
\tilde \Psi_{P^1}&=\psi _{0000}+\psi _{0011}-\psi _{0101}-\psi _{0110}-\psi _{1001}-\psi _{1010}+\psi _{1100}+\psi _{1111},\\
\tilde \Psi_{P^2}&=\psi _{0000}-\psi _{0011}+\psi _{0101}-\psi _{0110}-\psi _{1001}+\psi _{1010}-\psi _{1100}+\psi _{1111},\\
\tilde \Psi_{P^3}&=\psi _{0000}-\psi _{0011}-\psi _{0101}+\psi _{0110}+\psi _{1001}-\psi _{1010}-\psi _{1100}+\psi _{1111},\\
\tilde \Psi_{P^4}&=-\psi _{0000}-\psi _{0011}-\psi _{0101}-\psi _{0110}-\psi _{1001}-\psi _{1010}-\psi _{1100}-\psi _{1111}.
\eea
Note that the $SL(2)^{\times 4}$ invariant $\text{Tr}\tilde \Psi^2$ is a quadratic measure of four qubit entanglement\cite{verstraete}.

\subsection{The third split}

The previous two splits are related by triality of $\mathfrak{so}(4,4)$ and hence there must be one more inequivalent splitting of the algebra. Indeed this split is given by the $\pm 1$ eigenspaces of the involution
\beq
T\mapsto \eta_2 T^T \eta_2,
\eeq
where
\beq
\eta_2= \left(
\begin{array}{cc}
 \epsilon \otimes \epsilon &0\\
0 &  \epsilon \otimes \epsilon
\end{array}
\right),
\eeq
where $\epsilon=\left(
\begin{array}{cc}
 0 &1\\
-1 &  0
\end{array}
\right)$. The $-1$ eigenspace is 12 dimensional and spans $\mathfrak{sl}_2^{\times 4}$ algebra. The $+1$ eigenspace forms the representation $(2,2,2,2)$ under this. We do not need this split in the following hence we omit the explicit form of the generators.

\section{Explicit covariants for STU black holes}
\label{app:stumatrices}

Here we list the explicit forms of the covariants needed to calculate the $F$-invariant when the scalar asymptotics are set such that $X_i=0$ and $Y_i=1$.
\beq
R =\left(
\begin{array}{cccccc}
 \Sigma_1 & -\Xi_1 & 0 & 0 & 0 & 0 \\
 -\Xi_1 & -\Sigma_1 & 0 & 0 & 0 & 0 \\
 0 & 0 & \Sigma_2 & -\Xi_2 & 0 & 0 \\
 0 & 0 & -\Xi_2 & -\Sigma_2 & 0 & 0 \\
 0 & 0 & 0 & 0 & \Sigma_3 & -\Xi_3 \\
 0 & 0 & 0 & 0 & -\Xi_3 & -\Sigma_3 \\
\end{array}
\right),
\eeq
\beq
 K_{12} =\left(
\begin{array}{ccc}
 K_{12}^{(1)} & 0 & 0\\
0 &  K_{12}^{(2)} & 0\\
  0 & 0 &  K_{12}^{(3)}\\
  \end{array}
\right),
\eeq
where
\bea
 K_{12}^{(1)} &=\left(
\begin{array}{cc}
 -Q_2^2-Q_3^2-\left(P^1\right)^2-\left(P^4\right)^2 & -Q_4 P^1+Q_3 P^2+Q_2 P^3-Q_1 P^4 \\
 -Q_4 P^1+Q_3 P^2+Q_2 P^3-Q_1 P^4 & -Q_1^2-Q_4^2-\left(P^2\right)^2-\left(P^3\right)^2 \\
\end{array}
\right),\\
 K_{12}^{(2)} &=
 \left(
\begin{array}{cc}
 -Q_1^2-Q_3^2-\left(P^2\right)^2-\left(P^4\right)^2 & Q_3 P^1-Q_4 P^2+Q_1 P^3-Q_2 P^4 \\
 Q_3 P^1-Q_4 P^2+Q_1 P^3-Q_2 P^4 & -Q_2^2-Q_4^2-\left(P^1\right)^2-\left(P^3\right)^2 \\
\end{array}
\right),\\
 K_{12}^{(3)} &=
\left(
\begin{array}{cc}
 -Q_1^2-Q_2^2-\left(P^3\right)^2-\left(P^4\right)^2 & Q_2 P^1+Q_1 P^2-Q_4 P^3-Q_3 P^4 \\
 Q_2 P^1+Q_1 P^2-Q_4 P^3-Q_3 P^4 & -Q_3^2-Q_4^2-\left(P^1\right)^2-\left(P^2\right)^2 \\
\end{array}
\right)
\eea
\beq
 K_{11} =\left(
\begin{array}{ccc}
 K_{11}^{(1)} & 0 & 0\\
0 &  K_{11 }^{(2)} & 0\\
  0 & 0 &  K_{11}^{(3)}\\
  \end{array}
\right),
\eeq
where
\bea
 K_{11}^{(1)} &=
 \left(
\begin{array}{cc}
 Q_1 P^1-Q_2 P^2-Q_3 P^3+Q_4 P^4 & 2 \left(Q_1 Q_4+P^2 P^3\right) \\
 -2 \left(Q_2 Q_3+P^1 P^4\right) & -Q_1 P^1+Q_2 P^2+Q_3 P^3-Q_4 P^4 \\
\end{array}
\right),\\
 K_{11}^{(2)} &=
 \left(
\begin{array}{cc}
 -Q_1 P^1+Q_2 P^2-Q_3 P^3+Q_4 P^4 & 2 \left(Q_2 Q_4+P^1 P^3\right) \\
 -2 \left(Q_1 Q_3+P^2 P^4\right) & Q_1 P^1-Q_2 P^2+Q_3 P^3-Q_4 P^4 \\
\end{array}
\right),\\
 K_{11}^{(3)} &=
\left(
\begin{array}{cc}
 -Q_1 P^1-Q_2 P^2+Q_3 P^3+Q_4 P^4 & 2 \left(Q_3 Q_4+P^1 P^2\right) \\
 -2 \left(Q_1 Q_2+P^3 P^4\right) & Q_1 P^1+Q_2 P^2-Q_3 P^3-Q_4 P^4 \\
\end{array}
\right),
\eea
finaly
\beq
 K_{22} =\left(
\begin{array}{ccc}
 K_{22}^{(1)} & 0 & 0\\
0 &  K_{22 }^{(2)} & 0\\
  0 & 0 &  K_{22 }^{(3)}\\
  \end{array}
\right),
\eeq
where
\bea
 K_{22}^{(1)} &=
 \left(
\begin{array}{cc}
 -Q_1 P^1+Q_2 P^2+Q_3 P^3-Q_4 P^4 & 2 \left(Q_2 Q_3+P^1 P^4\right) \\
 -2 \left(Q_1 Q_4+P^2 P^3\right) & Q_1 P^1-Q_2 P^2-Q_3 P^3+Q_4 P^4 \\
\end{array}
\right),\\
 K_{22}^{(2)} &=
 \left(
\begin{array}{cc}
 Q_1 P^1-Q_2 P^2+Q_3 P^3-Q_4 P^4 & 2 \left(Q_1 Q_3+P^2 P^4\right) \\
 -2 \left(Q_2 Q_4+P^1 P^3\right) & -Q_1 P^1+Q_2 P^2-Q_3 P^3+Q_4 P^4 \\
\end{array}
\right),\\
 K_{22}^{(3)} &=
\left(
\begin{array}{cc}
 Q_1 P^1+Q_2 P^2-Q_3 P^3-Q_4 P^4 & 2 \left(Q_1 Q_2+P^3 P^4\right) \\
 -2 \left(Q_3 Q_4+P^1 P^2\right) & -Q_1 P^1-Q_2 P^2+Q_3 P^3+Q_4 P^4 \\
\end{array}
\right).
\eea

\bibliographystyle{utphys}
\bibliography{nonextremal}

\providecommand{\href}[2]{#2}\begingroup\raggedright\begin{thebibliography}{10}

\bibitem{compere1}
D.~D. Chow and G.~Comp{\`e}re, ``Seed for general rotating non-extremal black
  holes of $\mathcal{N}=8$ supergravity,''
  \href{http://dx.doi.org/10.1088/0264-9381/31/2/022001}{{\em Classical and
  quantum gravity} {\bfseries 31} no.~2, (2014) 022001},
  \href{http://arxiv.org/abs/1310.1925}{{\ttfamily arXiv:1310.1925 [hep-th]}}.

\bibitem{compere2}
D.~D. Chow and G.~Comp{\`e}re, ``Black holes in n= 8 supergravity from so (4,
  4) hidden symmetries,''
  \href{http://dx.doi.org/10.1103/PhysRevD.90.025029}{{\em Physical Review D}
  {\bfseries 90} no.~2, (2014) 025029},
  \href{http://arxiv.org/abs/1404.2602}{{\ttfamily arXiv:1404.2602 [hep-th]}}.

\bibitem{kallosh}
R.~Kallosh, N.~Sivanandam, and M.~Soroush, ``The non-bps black hole attractor
  equation,'' \href{http://dx.doi.org/10.1088/1126-6708/2006/03/060}{{\em
  Journal of High Energy Physics} {\bfseries 2006} no.~03, (2006) 060},
  \href{http://arxiv.org/abs/hep-th/0602005}{{\ttfamily arXiv:hep-th/0602005
  [hep-th]}}.

\bibitem{goldstein}
K.~Goldstein, N.~Iizuka, R.~P. Jena, and S.~P. Trivedi, ``Non-supersymmetric
  attractors,'' \href{http://dx.doi.org/10.1103/PhysRevD.72.124021}{{\em
  Physical Review D-Particles, Fields, Gravitation and Cosmology} {\bfseries
  72} no.~12, (2005) 124021\_1--124021\_24},
  \href{http://arxiv.org/abs/hep-th/0507096}{{\ttfamily arXiv:hep-th/0507096
  [hep-th]}}.

\bibitem{satokimura}
M.~Sato, T.~Kimura, {\em et~al.}, ``A classification of irreducible
  prehomogeneous vector spaces and their relative invariants,'' {\em Nagoya
  Math. J} {\bfseries 65} no.~1, (1977) 155.

\bibitem{andrianopoli2014extremal}
L.~Andrianopoli, A.~Gallerati, and M.~Trigiante, ``On extremal limits and
  duality orbits of stationary black holes,''
  \href{http://dx.doi.org/10.1007/JHEP01(2014)053}{{\em Journal of High Energy
  Physics} {\bfseries 2014} no.~1, (2014) 1--28},
  \href{http://arxiv.org/abs/1310.7886}{{\ttfamily arXiv:1310.7886 [hep-th]}}.

\bibitem{cirac}
W.~D{\"u}r, G.~Vidal, and J.~I. Cirac, ``Three qubits can be entangled in two
  inequivalent ways,'' \href{http://dx.doi.org/10.1103/PhysRevA.62.062314}{{\em
  Physical Review A} {\bfseries 62} no.~6, (2000) 062314},
  \href{http://arxiv.org/abs/quant-ph/0005115}{{\ttfamily
  arXiv:quant-ph/0005115 [quant-ph]}}.

\bibitem{verstraete}
F.~Verstraete, J.~Dehaene, B.~De~Moor, and H.~Verschelde, ``Four qubits can be
  entangled in nine different ways,''
  \href{http://dx.doi.org/10.1103/PhysRevA.65.052112}{{\em Physical Review A}
  {\bfseries 65} no.~5, (2002) 052112},
  \href{http://arxiv.org/abs/quant-ph/0109033}{{\ttfamily
  arXiv:quant-ph/0109033 [quant-ph]}}.

\bibitem{levvran1}
P.~L{\'e}vay and P.~Vrana, ``Three fermions with six single-particle states can
  be entangled in two inequivalent ways,''
  \href{http://dx.doi.org/10.1103/PhysRevA.78.022329}{{\em Physical Review A}
  {\bfseries 78} no.~2, (2008) 022329},
  \href{http://arxiv.org/abs/0806.4076}{{\ttfamily arXiv:0806.4076
  [quant-ph]}}.

\bibitem{duff2}
M.~J. Duff, ``String triality, black hole entropy, and cayley?s
  hyperdeterminant,'' \href{http://dx.doi.org/10.1103/PhysRevD.76.025017}{{\em
  Physical Review D} {\bfseries 76} no.~2, (2007) 025017},
  \href{http://arxiv.org/abs/hep-th/0601134}{{\ttfamily arXiv:hep-th/0601134
  [hep-th]}}.

\bibitem{CKW}
V.~Coffman, J.~Kundu, and W.~K. Wootters, ``Distributed entanglement,''
  \href{http://dx.doi.org/10.1103/PhysRevA.61.052306}{{\em Physical Review A}
  {\bfseries 61} no.~5, (2000) 052306},
  \href{http://arxiv.org/abs/quant-ph/9907047}{{\ttfamily
  arXiv:quant-ph/9907047 [quant-ph]}}.

\bibitem{bergshoeff2009generating}
E.~Bergshoeff, W.~Chemissany, A.~Ploegh, M.~Trigiante, and T.~Van~Riet,
  ``Generating geodesic flows and supergravity solutions,''
  \href{http://dx.doi.org/10.1016/j.nuclphysb.2008.10.023}{{\em Nuclear Physics
  B} {\bfseries 812} no.~3, (2009) 343--401},
  \href{http://arxiv.org/abs/0806.2310}{{\ttfamily arXiv:0806.2310 [hep-th]}}.

\bibitem{levay4qubit}
P.~L{\'e}vay, ``S t u black holes as four-qubit systems,''
  \href{http://dx.doi.org/10.1103/PhysRevD.82.026003}{{\em Physical Review D}
  {\bfseries 82} no.~2, (2010) 026003},
  \href{http://arxiv.org/abs/1004.3639}{{\ttfamily arXiv:1004.3639 [hep-th]}}.

\bibitem{duff1}
L.~Borsten, D.~Dahanayake, M.~J. Duff, A.~Marrani, and W.~Rubens, ``Four-qubit
  entanglement classification from string theory,''
  \href{http://dx.doi.org/10.1103/PhysRevLett.105.100507}{{\em Physical review
  letters} {\bfseries 105} no.~10, (2010) 100507},
  \href{http://arxiv.org/abs/1005.4915}{{\ttfamily arXiv:1005.4915 [hep-th]}}.

\bibitem{levay}
L.~Borsten, M.~Duff, and P.~Levay, ``The black-hole/qubit correspondence: an
  up-to-date review,''
  \href{http://dx.doi.org/10.1088/0264-9381/29/22/224008}{{\em Class. Quant.
  Grav.} {\bfseries 29} (2012) 224008},
  \href{http://arxiv.org/abs/1206.3166}{{\ttfamily arXiv:1206.3166 [hep-th]}}.

\bibitem{larsen}
M.~Cveti{\v{c}} and F.~Larsen, ``Black holes with intrinsic spin,''
  \href{http://dx.doi.org/10.1007/JHEP11(2014)033}{{\em Journal of High Energy
  Physics} {\bfseries 2014} no.~11, (2014) 1--32},
  \href{http://arxiv.org/abs/1406.4536}{{\ttfamily arXiv:1406.4536 [hep-th]}}.

\bibitem{cremmer}
E.~Cremmer, C.~Kounnas, A.~Van~Proeyen, J.~Derendinger, S.~Ferrara, B.~De~Wit,
  and L.~Girardello, ``Vector multiplets coupled to n= 2 supergravity:
  Super-higgs effect, flat potentials and geometric structure,''
  \href{http://dx.doi.org/10.1016/0550-3213(85)90488-2}{{\em Nuclear Physics B}
  {\bfseries 250} no.~1, (1985) 385--426}.

\bibitem{duff4}
M.~J. Duff, J.~T. Liu, and J.~Rahmfeld, ``Four-dimensional string/string/string
  triality,'' \href{http://dx.doi.org/10.1016/0550-3213(95)00555-2}{{\em
  Nuclear Physics B} {\bfseries 459} no.~1, (1996) 125--159},
  \href{http://arxiv.org/abs/hep-th/9508094}{{\ttfamily arXiv:hep-th/9508094
  [hep-th]}}.

\bibitem{cvetic}
M.~Cveti{\v{c}} and C.~M. Hull, ``Black holes and u-duality,''
  \href{http://dx.doi.org/10.1016/S0550-3213(96)00449-X}{{\em Nuclear Physics
  B} {\bfseries 480} no.~1, (1996) 296--316},
  \href{http://arxiv.org/abs/hep-th/9606193}{{\ttfamily arXiv:hep-th/9606193
  [hep-th]}}.

\bibitem{julia1}
E.~Cremmer and B.~Julia, ``The so (8) supergravity,''
  \href{http://dx.doi.org/10.1016/0550-3213(79)90331-6}{{\em Nuclear Physics B}
  {\bfseries 159} no.~1, (1979) 141--212}.

\bibitem{julia2}
E.~Cremmer and B.~Julia, ``The n= 8 supergravity theory. i. the lagrangian,''
  {\em Physics Letters B} {\bfseries 80} no.~1-2, (1978) 48--51.

\bibitem{breitenlohner}
P.~Breitenlohner, D.~Maison, and G.~Gibbons, ``4-dimensional black holes from
  kaluza-klein theories,'' \href{http://dx.doi.org/10.1007/BF01217967}{{\em
  Communications in Mathematical Physics} {\bfseries 120} no.~2, (1988)
  295--333}.

\bibitem{carbone}
L.~Carbone, S.~Murray, and H.~Sati, ``Integral group actions on symmetric
  spaces and discrete duality symmetries of supergravity theories,''
  \href{http://arxiv.org/abs/1407.3370}{{\ttfamily arXiv:1407.3370 [hep-th]}}.

\bibitem{levsar1}
G.~S{\'a}rosi and P.~L{\'e}vay, ``Entanglement in fermionic fock space,''
  \href{http://dx.doi.org/10.1088/1751-8113/47/11/115304}{{\em Journal of
  Physics A: Mathematical and Theoretical} {\bfseries 47} no.~11, (2014)
  115304}, \href{http://arxiv.org/abs/1309.4300}{{\ttfamily arXiv:1309.4300
  [quant-ph]}}.

\bibitem{levsar2}
G.~S{\'a}rosi and P.~L{\'e}vay, ``Entanglement classification of three fermions
  with up to nine single-particle states,''
  \href{http://dx.doi.org/10.1103/PhysRevA.89.042310}{{\em Physical Review A}
  {\bfseries 89} no.~4, (2014) 042310},
  \href{http://arxiv.org/abs/1312.2786}{{\ttfamily arXiv:1312.2786
  [quant-ph]}}.

\bibitem{levsar3}
G.~S{\'a}rosi and P.~L{\'e}vay, ``Coffman-kundu-wootters inequality for
  fermions,'' \href{http://dx.doi.org/10.1103/PhysRevA.90.052303}{{\em Physical
  Review A} {\bfseries 90} no.~5, (2014) 052303},
  \href{http://arxiv.org/abs/1408.6735}{{\ttfamily arXiv:1408.6735
  [quant-ph]}}.

\bibitem{rasheed1995rotating}
D.~Rasheed, ``The rotating dyonic black holes of kaluza-klein theory,''
  \href{http://dx.doi.org/10.1016/0550-3213(95)00396-A}{{\em Nuclear Physics B}
  {\bfseries 454} no.~1, (1995) 379--401},
  \href{http://arxiv.org/abs/hep-th/9505038}{{\ttfamily arXiv:hep-th/9505038
  [hep-th]}}.

\bibitem{lozano2000general}
E.~Lozano-Tellechea and T.~Ort{\i}n, ``The general, duality-invariant family of
  non-bps black-hole solutions of n= 4, d= 4 supergravity,''
  \href{http://dx.doi.org/10.1016/S0550-3213(99)00762-2}{{\em Nuclear Physics
  B} {\bfseries 569} no.~1, (2000) 435--450},
  \href{http://arxiv.org/abs/hep-th/9910020}{{\ttfamily arXiv:hep-th/9910020
  [hep-th]}}.

\bibitem{cvetivc1996entropy}
M.~Cveti{\v{c}} and D.~Youm, ``Entropy of nonextreme charged rotating black
  holes in string theory,''
  \href{http://dx.doi.org/10.1103/PhysRevD.54.2612}{{\em Physical Review D}
  {\bfseries 54} no.~4, (1996) 2612},
  \href{http://arxiv.org/abs/hep-th/9603147}{{\ttfamily arXiv:hep-th/9603147
  [hep-th]}}.

\bibitem{kalloshlinde}
R.~Kallosh and A.~Linde, ``Strings, black holes, and quantum information,''
  \href{http://dx.doi.org/10.1103/PhysRevD.73.104033}{{\em Physical Review D}
  {\bfseries 73} no.~10, (2006) 104033},
  \href{http://arxiv.org/abs/hep-th/0602061}{{\ttfamily arXiv:hep-th/0602061
  [hep-th]}}.

\bibitem{duff5}
M.~J. Duff and S.~Ferrara, ``E 7 and the tripartite entanglement of seven
  qubits,'' \href{http://dx.doi.org/10.1103/PhysRevD.76.025018}{{\em Physical
  Review D} {\bfseries 76} no.~2, (2007) 025018},
  \href{http://arxiv.org/abs/hep-th/0609227}{{\ttfamily arXiv:hep-th/0609227
  [hep-th]}}.

\bibitem{lekeu}
G.~Comp{\`e}re and V.~Lekeu, ``$e_{7(7)}$ invariant non-extremal entropy,''
  \href{http://arxiv.org/abs/1510.03582}{{\ttfamily arXiv:1510.03582
  [hep-th]}}.

\bibitem{witten}
J.~Maldacena, A.~Strominger, and E.~Witten, ``Black hole entropy in m-theory,''
  \href{http://dx.doi.org/10.1088/1126-6708/1997/12/002}{{\em Journal of High
  Energy Physics} {\bfseries 1997} no.~12, (1997) 002},
  \href{http://arxiv.org/abs/hep-th/9711053}{{\ttfamily arXiv:hep-th/9711053
  [hep-th]}}.

\bibitem{horowitz1}
G.~T. Horowitz, D.~A. Lowe, and J.~M. Maldacena, ``Statistical entropy of
  nonextremal four-dimensional black holes and u duality,''
  \href{http://dx.doi.org/10.1103/PhysRevLett.77.430}{{\em Physical Review
  Letters} {\bfseries 77} no.~3, (1996) 430},
  \href{http://arxiv.org/abs/hep-th/9603195}{{\ttfamily arXiv:hep-th/9603195
  [hep-th]}}.

\bibitem{horowitz2}
R.~Emparan and G.~T. Horowitz, ``Microstates of a neutral black hole in m
  theory,'' \href{http://dx.doi.org/10.1103/PhysRevLett.97.141601}{{\em
  Physical review letters} {\bfseries 97} no.~14, (2006) 141601},
  \href{http://arxiv.org/abs/hep-th/0607023}{{\ttfamily arXiv:hep-th/0607023
  [hep-th]}}.

\bibitem{horowitz3}
G.~T. Horowitz and M.~M. Roberts, ``Counting the microstates of a kerr black
  hole in m theory,''
  \href{http://dx.doi.org/10.1103/PhysRevLett.99.221601}{{\em Physical review
  letters} {\bfseries 99} no.~22, (2007) 221601},
  \href{http://arxiv.org/abs/0708.1346}{{\ttfamily arXiv:0708.1346 [hep-th]}}.

\bibitem{alejandra}
A.~Castro, A.~Maloney, and A.~Strominger, ``Hidden conformal symmetry of the
  kerr black hole,'' \href{http://dx.doi.org/10.1103/PhysRevD.82.024008}{{\em
  Physical Review D} {\bfseries 82} no.~2, (2010) 024008},
  \href{http://arxiv.org/abs/1004.0996}{{\ttfamily arXiv:1004.0996 [hep-th]}}.

\bibitem{strominger}
D.~Gaiotto, A.~Strominger, and X.~Yin, ``New connections between 4d and 5d
  black holes,'' \href{http://dx.doi.org/10.1088/1126-6708/2006/02/024}{{\em
  Journal of High Energy Physics} {\bfseries 2006} no.~02, (2006) 024},
  \href{http://arxiv.org/abs/hep-th/0503217}{{\ttfamily arXiv:hep-th/0503217
  [hep-th]}}.

\bibitem{krutelevich}
S.~Krutelevich, ``Jordan algebras, exceptional groups, and bhargava
  composition,'' \href{http://dx.doi.org/10.1016/j.jalgebra.2007.02.060}{{\em
  Journal of algebra} {\bfseries 314} no.~2, (2007) 924--977}.

\bibitem{duff3}
L.~Borsten, D.~Dahanayake, M.~Duff, and W.~Rubens, ``Black holes admitting a
  freudenthal dual,'' \href{http://dx.doi.org/10.1103/PhysRevD.80.026003}{{\em
  Physical Review D} {\bfseries 80} no.~2, (2009) 026003},
  \href{http://arxiv.org/abs/0903.5517}{{\ttfamily arXiv:0903.5517 [hep-th]}}.

\bibitem{virmani}
D.~Katsimpouri, A.~Kleinschmidt, and A.~Virmani, ``An inverse scattering
  formalism for stu supergravity,''
  \href{http://dx.doi.org/10.1007/JHEP03(2014)101}{{\em Journal of High Energy
  Physics} {\bfseries 2014} no.~3, (2014) 1--27},
  \href{http://arxiv.org/abs/1311.7018}{{\ttfamily arXiv:1311.7018 [hep-th]}}.

\bibitem{vernocke}
S.~Banerjee, B.~D. Chowdhury, B.~Vercnocke, and A.~Virmani,
  ``Non-supersymmetric microstates of the msw system,''
  \href{http://dx.doi.org/10.1007/JHEP05(2014)011}{{\em Journal of High Energy
  Physics} {\bfseries 2014} no.~5, (2014) 1--34},
  \href{http://arxiv.org/abs/1402.4212}{{\ttfamily arXiv:1402.4212 [hep-th]}}.

\end{thebibliography}\endgroup

\end{document}